\documentclass{article}
\usepackage{graphicx} 
\usepackage{authblk}
\usepackage[style=nature,backend=biber,maxnames=99,autocite=superscript]{biblatex}
\usepackage{booktabs}
\usepackage[section]{placeins}
\usepackage[inkscapelatex=false]{svg}
\usepackage[margin=1in]{geometry}
\usepackage{newfloat}
\usepackage{amsmath}
\usepackage{appendix}
\usepackage{lineno}
\DeclareFloatingEnvironment[name={Supplementary Figure}]{suppfigure}
\title{Validity in machine learning for extreme event attribution}
\author[1]{Cassandra C. Chou\thanks{Corresponding author. Email: cassiechou@g.harvard.edu}}
\author[1]{Scott L. Zeger}
\author[2]{Benjamin Q. Huynh}
\affil[1]{Department of Biostatistics, Johns Hopkins University}
\affil[2]{Department of Environmental Health \& Engineering, Johns Hopkins University}

\date{}

\begin{document}

\maketitle
\begin{abstract}
Extreme event attribution (EEA), which assesses the extent to which disasters are caused by climate change, is crucial for informing climate policy and legal proceedings. Machine learning is increasingly used for EEA by modeling rare weather events too complex or computationally intensive for traditional methods. However, its validity remains unclear, as machine learning applications are criticized for bias and lack of robustness. Here we evaluate machine learning for EEA using California wildfire data from 2003-2020. We identify three major threats to validity: (1) individual event attribution estimates are highly sensitive to algorithmic design choices; (2) common performance metrics like Brier score are not strongly correlated with attribution error, facilitating suboptimal model selection; and (3) distribution shift across climate scenarios substantially degrades predictive performance. We propose a more robust attribution analysis using aggregate estimates and additional evaluation metrics for predictive performance and distribution shift.

\end{abstract}

\section{Introduction}\label{intro}

Extreme weather events are increasing in frequency and magnitude, driven by human-induced changes in climate.\autocite{noauthor_chapter_nodate} To quantify this increase, extreme event attribution (EEA) seeks to evaluate the impact of climate change on the likelihood of extreme weather events. Most EEA approaches compare the probability of extreme events under historically observed conditions with the probability of events under counterfactual pre-industrial conditions representing a world without human-induced climate change \autocite{stott_attribution_2016, otto_attribution_2017, perkins-kirkpatrick_attribution_2022}. 

Courts are using attribution studies to determine liability for the damages caused by extreme events. \autocite{lloyd_climate_2021, burger_law_2020, marjanac_extreme_2018, adam_climate_2011} EEA can also be used to compare current scenarios to projected climate scenarios, predicting the impact of future climate change to inform adaptation planning. \autocite{clarke_inventories_2021,thompson_ethical_2015}

Most attribution analyses use global climate models to perform physics-based simulations of the climate system under different warming scenarios to assess human influence on extreme weather events. In weather forecasting more broadly, AI and machine learning techniques are being adopted due to superior predictive performance and reduced computational costs. \autocite{bi_accurate_2023,liu_evaluation_2024,charlton-perez_ai_2024,kurth_fourcastnet_nodate,lam_learning_2023,price_probabilistic_2025,vaughan_aardvark_2024,camps-valls_artificial_2025}  Similarly, machine learning has been used to facilitate EEA for phenomena such as wildfires, heat waves, and tropical cyclones, enabling attribution estimates that would otherwise be too complex or computationally expensive to model through traditional approaches.\autocite{trok_machine_2024,jimenez-esteve_ai-driven_2024,loridan_reask_2025,brown_climate_2023}

The validity of machine learning for EEA remains unclear. Machine learning tools have performed poorly in high-stakes applications such as criminal justice or healthcare, raising questions of bias and robustness.\autocite{buolamwini_gender_2018,dressel_accuracy_2018,corbett-davies_algorithmic_2017,obermeyer_dissecting_2019,wong_external_2021} Therefore, it is crucial to assess the extent to which machine learning can introduce uncertainty, bias, or other threats to validity in EEA.\autocite{behzad_reconciling_2025,dai_be_2025,angelopoulos_prediction-powered_2023,ovadia_can_2019}

Here we analyze daily California wildfire data from 2003-2020 -- which includes 17,910 fire days across varying locations -- evaluating the conditions under which machine learning can be used to produce valid attribution estimates. We investigate three threats to validity: \begin{enumerate}
    \item Is attribution of individual events via machine learning subject to model multiplicity, where predictive models may exhibit different individual-level estimates despite similar aggregate predictive performance? \autocite{dai_be_2025}
    \item Does selection of an “optimal”  model or algorithm using performance metrics assure that the method is optimal for making attribution estimates such as fraction of attributable risk (FAR) or risk ratio (RR), the scientific focus?
    \item Do counterfactual climate scenarios represent distribution shifts in predictor variables that can degrade predictive performance, inhibit model selection, and bias FAR estimation?
\end{enumerate} 

We first replicate an existing approach to using machine learning for EEA of California wildfires.\autocite{brown_climate_2023} A machine learning model is trained on historical environmental data to predict extreme daily wildfire growth -- defined as wildfire growth over 10,000 acres in a single day -- then applied to a counterfactual dataset generated using global climate models and representing different climate scenarios. The ratio of predicted values under the counterfactual data to the observed values in the observed data represent the attribution estimate, which is the measure of the effect of climate change on extreme wildfire risk (Figure \ref{fig:fig1}). This approach is a form of \textit{storyline attribution}, where computational models are used to assess how historical events may have played out differently under different climate conditions. To evaluate model multiplicity, we repeat this process with competing machine learning models and assess the variability in attribution estimates. We evaluate the robustness of both the individual-day attribution estimates and aggregate attribution estimates -- defined as the planetary level impact of climate change -- across climate scenarios.

Next, we conduct simulation experiments in which true probabilities of extreme wildfires and the attribution effect are known. We generate new datasets, train machine learning and post-hoc calibration models to make attribution estimates, and assess which internal performance metrics are most predictive of an accurate attribution estimate. We quantify the degree to which attribution estimates are affected by stronger distribution shifts, represented by the severity of warming. We analyze two counterfactual climate scenarios: (1) a standard pre-industrial scenario and (2) a worst-case SSP5-8.5 end-of-century scenario. Finally, we present recommendations for more valid ML-based event attribution based on our results, calculating a novel attribution estimate.

\section{Results}\label{results}
\subsection{Event attribution estimates are highly sensitive to model choice}
When comparing the present-day to the pre-industrial climate scenario, models with similar predictive performance produce attribution estimates with conflicting implications (presented as $RR_i$ estimates that cross 1 in Figure \ref{fig:fig2}). This occurs in 66.58\% of extreme fire growth days (Figure \ref{fig:fig1}a) despite areas under the receiver operating characteristic curve (AUC) ranging from 0.86 to 0.89 (Figure \ref{fig:fig1}b), indicating fundamental disagreement on the directional impact of climate change on a wildfire day. The median range factor of RR estimates, or the ratio of the maximum RR estimate to the minimum RR estimate across the six machine learning models, for an individual day was 4.66 (IQR: [3.14, 7.96]). This means that on a typical day, the largest risk ratio estimate was 4.66 times larger than the lowest estimate.

These results are qualitatively consistent across different climate and fire day scenarios. When considering all fire days and not only days with extreme fire growth, models produced estimates with conflicting implications in 62.19\% of all days (median range factor = 5.51 [3.15, 11.26]). When comparing present-day to a worst-case future climate scenario (SSP5-8.5, end-of-century) instead of pre-industrial emissions levels, models produced estimates with conflicting implications in 18.42\% of extreme fire growth days (median range factor = 3.36 [2.49, 5.56]). When considering all fire days, models produced estimates with conflicting implications in 56.56\% of observations (median range factor = 6.91 [3.66, 17.60]).

Aggregate estimates are more robust than individual-day estimates, with no models producing conflicting implications in either climate scenario comparison for aggregate RR estimates. Similarly, aggregate RR estimates are smaller in range factor than the median range factor estimates for individual days, with range factors of 1.18 and 1.65 for pre-industrial and end-of-century comparisons, respectively. (Extended Data Table \ref{tab:supp1} and Table \ref{tab:tab1}).

\subsection{Traditional performance metrics lead to inaccurate attribution estimates}
Traditional predictive performance metrics such as AUC and Brier skill score showed weak to moderate correlations with log risk ratio error in simulation analyses. Each correlation is taken across the absolute log risk ratio error and performance estimates calculated on the same simulated dataset from twenty five combinations of machine learning model and calibration methods. (Figure \ref{fig:fig4}) Median correlation between AUC and absolute log risk ratio error was low (median r = 0.19 (IQR: [0.10, 0.28]). By comparison, median correlation between Brier skill score and absolute log risk ratio error was significantly higher (median r = 0.43 [0.32, 0.57]; p $<$ 0.0001). Further, mean calibration error has a much stronger correlation with log risk ratio error than Brier skill score (median r = 0.76 [0.52, 0.87]; p $<$ 0.0001). (Figure  \ref{fig:fig4}a, Extended Data Table \ref{tab:supp2})

These results do not hold in the SSP5-8.5 end-of-century scenario comparison, where the magnitude of distribution shift is larger, arising from a vastly different climate. While AUC and Brier skill score similarly showed weak correlations with absolute log risk ratio error (median r = 0.36 [0.19, 0.50] and r = 0.32 [0.18, 0.49]; AUC vs. BSS, p = 0.98), the correlation between mean calibration error and absolute log risk ratio error remains weak (median r = 0.36 [0.22, 0.48]; AUC vs. MCE, p = 0.73; BSS vs. MCE, p  = 0.63). (Table \ref{tab:supp2})

The regret analysis demonstrates that when selecting the top 5 best performing models for each metric and taking the worst estimate, no metric results in significantly more accurate estimates (Figure \ref{fig:fig4}b). Regret is calculated as the highest log risk ratio error across the five machine learning methods, excluding the ensemble, and calibration method combinations with the best performance for each of the 500 simulations.

Using the estimates of mean calibration error, Brier score, and Brier skill score from our models trained on the observed present-day data, we find that all five machine learning methods are appropriate for this context. Using the ensemble estimates with all 5 machine learning methods, we estimate that 17.5\% (95\% Confidence Interval: (9.7\%, 27.0\%)) of the current risk of extreme wildfire growth is attributable to human-induced climate change. We estimate that 52.3\% (32.4\%, 59.1\%) of the SSP5-8.5 end-of-century risk of extreme wildfire growth is attributable to human-induced climate change that occurs from the present-day (\ref{tab:tab1}). 

\subsection{Models degrade under climate change-induced distribution shift}
The distribution shift from present-day to SSP5-8.5, end of century is much larger than the shift from pre-industrial to present-day, as the modeled temperature differences are much larger. Proxy-A distance, a metric for distribution shift, is 0.405 and 1.378 for the pre-industrial and SSP5-8.5, end-of-century scenarios respectively. 

The temperature subanalysis -- where we subgrouped our data by temperature and calculated mean calibration error on these subgroups -- demonstrates that predictive performance decreases with temperature in historically observed data. Average mean calibration error increased from 0.000821 at the lowest temperatures to 0.0288 at the highest temperatures, corresponding to a relative increase of 3408\% (Figure \ref{fig:fig5}a).  

The principal components analysis of covariates under different climate scenarios further illustrates that the shift in principal components from present-day to pre-industrial is smaller than from present-day to SSP5-8.5, end of century (Figure \ref{fig:fig5}c).

Similarly, out-of-sample predictive performance decreases with distribution shift; out-of-sample mean calibration error increases from 0.00126 (IQR: (0.000530, 0.00241)) in the pre-industrial scenario to 0.00632 (0.00281, 0.0118) in the SSP5-8.5, end-of-century scenario. Additionally, mean calibration error correlation with absolute log risk ratio error is weaker under stronger distribution shift (Table \ref{tab:supp2}).

\subsection{Addressing distribution shift and model error}
Weighted Prediction-Powered Inference (PPI) leads to more accurate attribution estimates in comparison to plug-in and unweighted PPI methods in simulation analyses. Coverage of the true attribution estimate in the pre-industrial climate scenario is highest for weighted PPI (Coverage = 0.68), followed by plug-in (0.25), and lowest for unweighted PPI (0.02). Similarly, the log risk ratio error for weighted PPI estimates was lowest in magnitude (Median = -0.013, IQR: [-0.038, 0.011]) in comparison to plug-in (-0.157 [-0.194, -0.121]) and unweighted PPI (-0.042 [-0.132, 0.060]). More concretely, the median absolute log RR error is 5.94 higher for plug-in estimates in comparison to weighted PPI, and 3.74 for unweighted PPI estimates in comparison to weighted PPI. These results hold when varying model performance. 

These results also hold in the SSP5-8.5, end-of-century scenario, despite the variance in the log risk ratio error being larger across all methods. Coverage is highest for weighted PPI (0.6), followed by plug-in (0.31), lowest for unweighted PPI (0.11). Log risk ratio error for weighted PPI was lowest in magnitude (0.016 [-0.114, 0.144]), followed by unweighted PPI (0.042 [-0.245, 0.383]), followed lastly by plug-in (-0.16 [-0.257, -0.062]). Similarly, the median absolute log RR error is 1.31 higher for plug-in estimates in comparison to weighted PPI, and 2.45 for unweighted PPI estimates in comparison to weighted PPI (Figure \ref{fig:fig3}). 

\section{Discussion}

In this study, we identified threats to validity in using machine learning for EEA: model multiplicity, error introduced by machine learning, misleading performance metrics, and distribution shift. Although these findings indicate that current applications of machine learning to EEA may be vulnerable to these threats, we find that it may be possible to mitigate them. Specifically, we recommend the following techniques: (1) comparing multiple models to quantify epistemic uncertainty; (2) prioritizing calibration over AUC as a performance metric; (3) using diagnostics to measure distribution shift and weighting to adjust for it; (4) using prediction-powered inference to correct for machine learning error and distribution shift error.

One remaining concern not addressed by the above recommendations is our finding that individual-day attribution estimates are highly sensitive to model design choices. As individual probabilities of events are underdetermined by the data, it is not possible to guarantee calibrated probabilities for individual observations using this particular approach of machine learning for EEA, where each machine learning output corresponds to a probability estimate for a single extreme event. It is for this reason that we recommend using aggregate attribution estimates over multiple extreme events rather than individual event attribution estimates – we found aggregate estimates to be far more robust to multiplicity, with no conflicting signs and only modest differences between models. This corresponds to prior findings that raw estimates of probability are not fully appropriate for impact attribution. \autocite{perkins-kirkpatrick_attribution_2022}

In a worst scenario, if volatile, inconsistent individual probability estimates are overinterpreted, this could lead to misuse by motivated actors. For example, this high sensitivity and inconsistency could be exploited by motivated actors to obtain desired results, such as in a climate litigation setting where plaintiffs and defendants may introduce competing model estimates to determine liability for a single event.

Existing approaches such as post hoc calibration, ensemble learning, or reconciliation are likely insufficient to resolve multiplicity in machine learning for EEA. Post hoc calibration works by training a calibration model on top of model estimates to generate more accurate probability estimates; however,  extreme events are, by definition, rare, yielding a low effective sample size and reducing the effectiveness of calibration. Ensemble learning and reconciliation both follow the principle that single-model estimates may be improved by aggregating predictions from multiple models. However, such approaches only work properly in ``good faith" settings where model developers have the goal of finding the most accurate model as ensembled and reconciled models are themselves additional candidates in the model space. In an adversarial setting like climate litigation, motivated actors can still selectively construct ensembles by only including models advantageous for their needs.\autocite{behzad_reconciling_2025,roth_reconciling_2023} Nevertheless, including multiple models is a straightforward method of assessing variation due to model choice, and we therefore recommend reporting results from different models.

We recommend using mean calibration error as an additional predictive performance metric because it more directly relates to the attribution estimates, FAR and RR. Attribution error arises from poorly calibrated probability estimates, but metrics like AUC measure discriminative capacity, not calibration. Brier score and Brier skill score decompose into both discrimination and calibration components, indicating they are not entirely suitable for estimating attribution error. Still, our findings in the regret analysis indicate that mean calibration error is necessary but not sufficient for identifying the most accurate model. Brier score or Brier skill score can be used in support.

We found that optimal model selection is impaired by distribution shift. Mean calibration was strongly correlated with attribution error in a pre-industrial scenario, but not in an end-of-century SSP5-8.5 scenario. A worst-case scenario like SSP5-8.5 is more drastic of a change than a pre-industrial climate scenario, so the extent of distribution shift is greater. 

To assess distribution shift in EEA, we recommend two diagnostics: propensity models and temperature subgroup analyses. Propensity models can be used to assess distribution shift by training machine learning models to discriminate between two different climate scenarios. If propensity scores have low overlap, then the two scenarios are highly separable, indicating severe distribution shift (Figure 1). This use of propensity scores as a distance metric is also represented by proxy-A distance.\autocite{ben-david_analysis_2006} In our temperature subgroup analyses, we found that predictive performance decreased as temperature rises, suggesting transporting models in our context works well when predicting for a cooler climate (i.e., a pre-industrial scenario) compared to a warmer climate (i.e., a future climate scenario). Although we found weighted PPI to be more effective in recovering the true attribution effect compared to unweighted estimates, overall estimates were noisier under the strong distribution shift of the end-of-century scenario compared to a weaker distribution shift of the pre-industrial scenario. Machine learning for EEA is therefore more likely to be robust when the extent of distribution shift is low and subgroup analyses suggest transportability.

Our work has several limitations. First, we only consider wildfires – we do not assess other types of extreme events. As such, we identify issues and solutions related broadly to concepts and metrics for machine learning methods, rather than seeking the best specific models, which will vary depending on context. Further, our dataset is limited to California wildfire days from 2003-2020, where only 2\% of observations were considered extreme wildfire days – the relative rarity of extreme events may limit the effectiveness of more data-intensive procedures such as post hoc calibration. Given the rare nature of extreme events, machine learning analyses should not assume large-scale availability of extreme event data. Another limitation of our work is the dependence of our data within wildfires; because we use daily wildfire data, many of our positive cases represent different days of the same wildfire. We mitigated this using temporal cross validation, where we split our data into three-year time intervals to contain the same wildfires in the same fold, but some wildfires may extend across time intervals. Additionally, we only considered five different machine learning models as an illustrative point, which does not provide a full estimate of model multiplicity. If we were to consider the full range of possible models, architectures, hyperparameters, and predictors, the true extent of multiplicity would be higher than what we estimated. Finally, we only examine the uncertainty arising from introducing machine learning to EEA – there are other sources of uncertainty intrinsic to EEA that are out of scope for this analysis.

Our work expands previous work from Brown et al.\autocite{brown_climate_2023}, which used machine learning to estimate attribution comparing the pre-industrial to present-day and various future climate scenarios using wildfire data. We also expand on previous work that evaluates methods of assessing the accuracy of FAR estimates in simulations\autocite{lott_evaluating_2016}. We draw from previous work that uses artificial intelligence for EEA more broadly, with prior studies including machine learning for attribution for heat waves\autocite{trok_machine_2024,callahan_increasing_2026}, tropical cyclones\autocite{loridan_reask_2025}, and weather forecasting for extreme events and EEA.\autocite{jimenez-esteve_ai-driven_2024} We also draw on previous work that uses statistical methods for EEA, including a prior study using extreme value theory to address modeling uncertainties.\autocite{naveau_statistical_2020} Relevant previous work also includes applications of AI and machine learning techniques for weather forecasting.\autocite{bi_accurate_2023, liu_evaluation_2024, charlton-perez_ai_2024, kurth_fourcastnet_nodate, lam_learning_2023, price_probabilistic_2025, vaughan_aardvark_2024, camps-valls_artificial_2025} Our work contributes to the literature by identifying challenges to validity from using machine learning for EEA.

\section{Conclusion}
Machine learning is increasingly being used to model extreme weather phenomena, and it is likely this trend will extend to extreme event attribution. Given the potential for machine learning applications in high-stakes scenarios to yield catastrophic results, it is imperative to establish best practices and validity conditions. Our findings and recommendations can be used to improve model robustness, but are largely technical in nature. Given attribution science’s rising role in climate litigation, there will likely be sociolegal issues arising from the use of AI in EEA, ranging from extant governance concerns such as reproducibility and transparency to more modern ones, such as adversarial uses of AI or opaque black-box decision-making. As AI advances, it is a critical priority to determine how to responsibly use such models for effective climate research and policy.

\newpage

\section{Figures}
\begin{figure}[ht]
    \centering
    \includegraphics[width=1\linewidth]{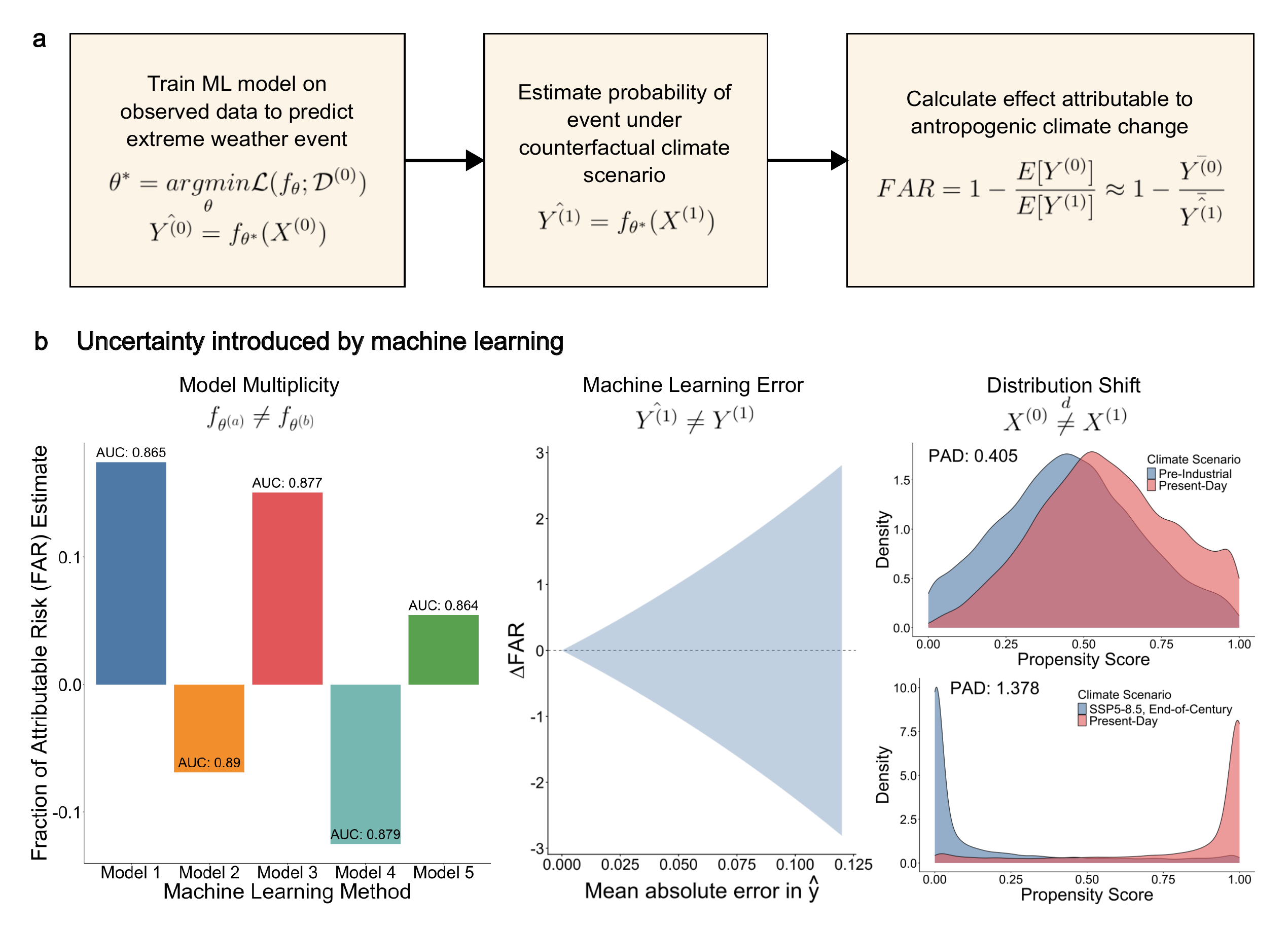}
    \caption[Machine learning framework for extreme event attribution, including challenges to validity introduced by machine learning.]{\textbf{Machine learning framework for extreme event attribution, including challenges to validity introduced by machine learning.} \textbf{a}, Overview of machine learning methodology to conduct extreme event attribution. \textbf{b}, Challenges to validity introduced by machine learning: error in machine learning estimates of event probability leads to higher error in fraction of attributable risk (FAR) estimates; propensity score separability indicates distribution shift under different climate scenarios; FAR estimates vary widely for an example extreme fire day using five different machine learning models with similar predictive performance, demonstrating model multiplicity.}
    \label{fig:fig1}
\end{figure}

\newpage
\begin{figure}[ht]
    \centering
    \includegraphics[width=1\linewidth]{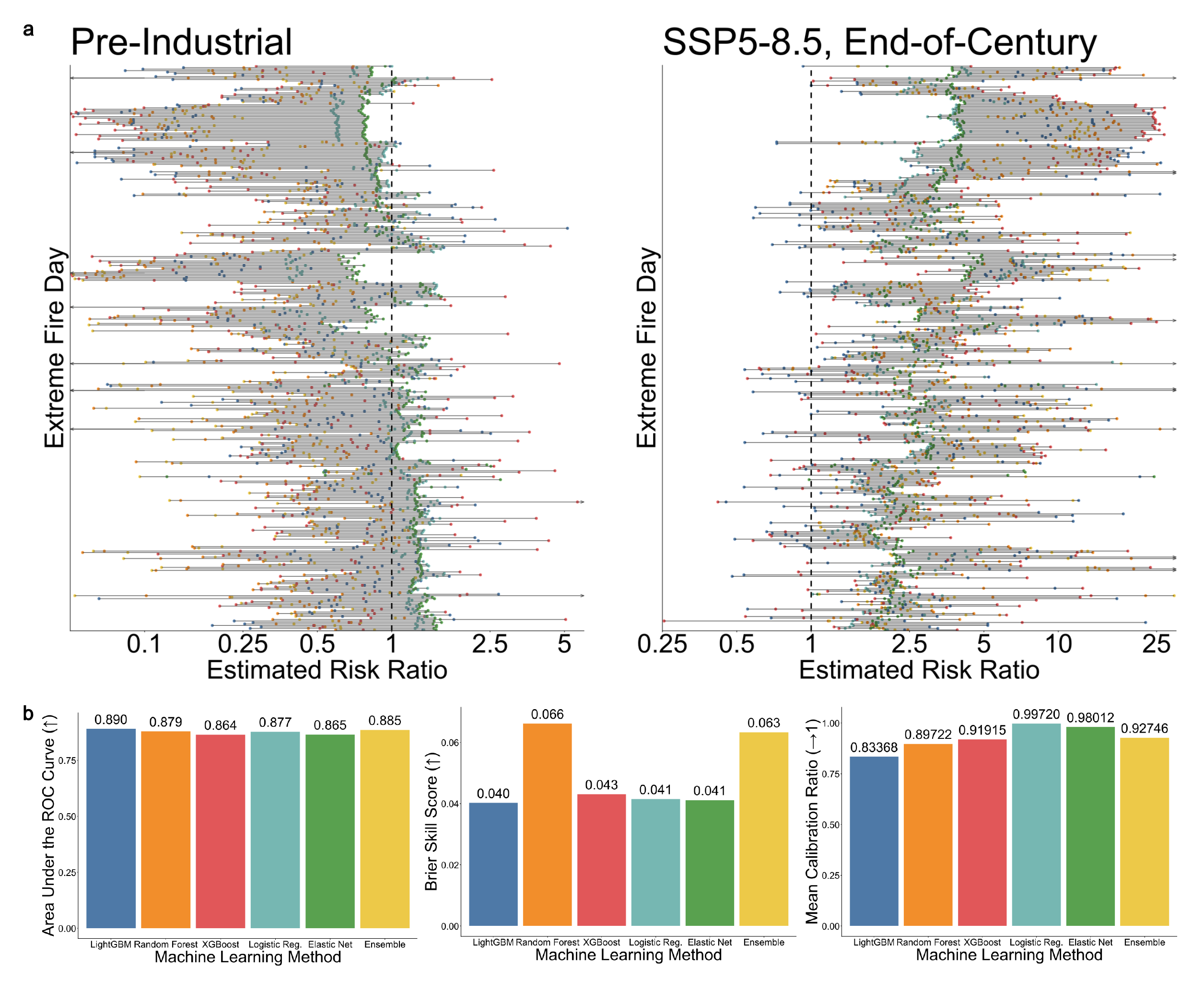}
    \caption[Sensitivity of individual-event estimates to different model specifications, indicating model multiplicity.]{\textbf{Sensitivity of individual-event estimates to different model specifications, indicating model multiplicity.} \textbf{a}, Estimated FAR for 380 extreme fire days in a pre-industrial scenario and an end-of-century SSP5-8.5 scenario: horizontal lines indicate range of estimates across different machine learning models; arrows represent values outside the frame; horizontal lines crossing zero model disagreement on the directional impact of climate change. \textbf{b}, Performance metrics to evaluate machine learning models.}
    \label{fig:fig2}
\end{figure}

\newpage
\begin{figure}[ht]
    \centering
    \includegraphics[width=1\linewidth]{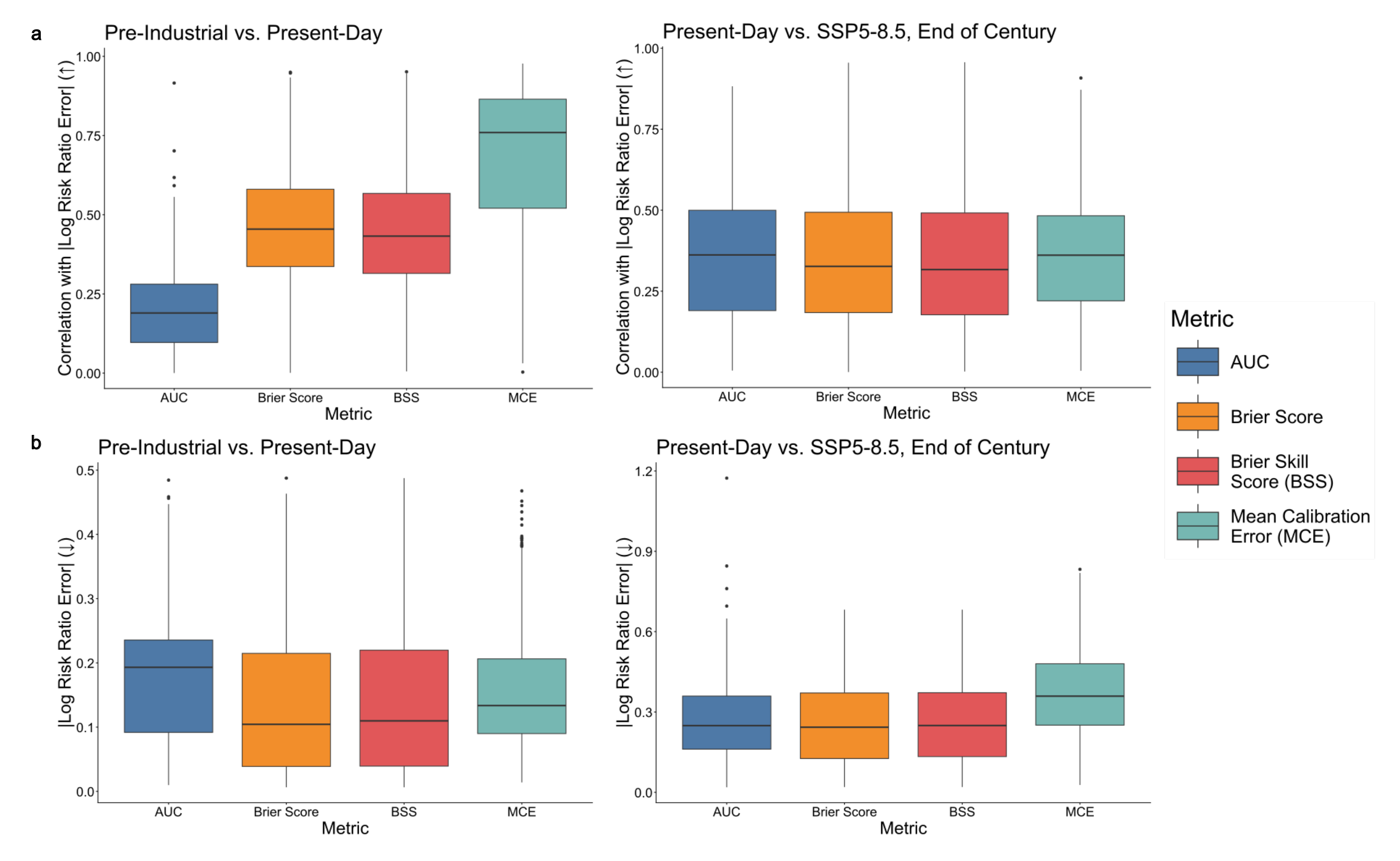}
    \caption[Relationship between internal predictive performance metrics and accuracy of the fraction of attributable risk estimate, modeled across 500 simulations.]{\textbf{Relationship between internal predictive performance metrics and accuracy of the fraction of attributable risk estimate, modeled across 500 simulations.} \textbf{a}, Distribution of correlation between internal predictive performance metrics and accuracy of risk ratios within a simulation. Correlations taken across 25 machine learning and calibration method combinations.  Distribution taken across 500 simulations.\textbf{b}, Regret analysis; distribution of worst absolute log risk ratio error across top 5 best performing models for each metric. Distribution taken across 500 simulations.}
    \label{fig:fig4}
\end{figure}

\newpage
\begin{figure}[ht]
    \centering
    \includegraphics[width=1\linewidth]{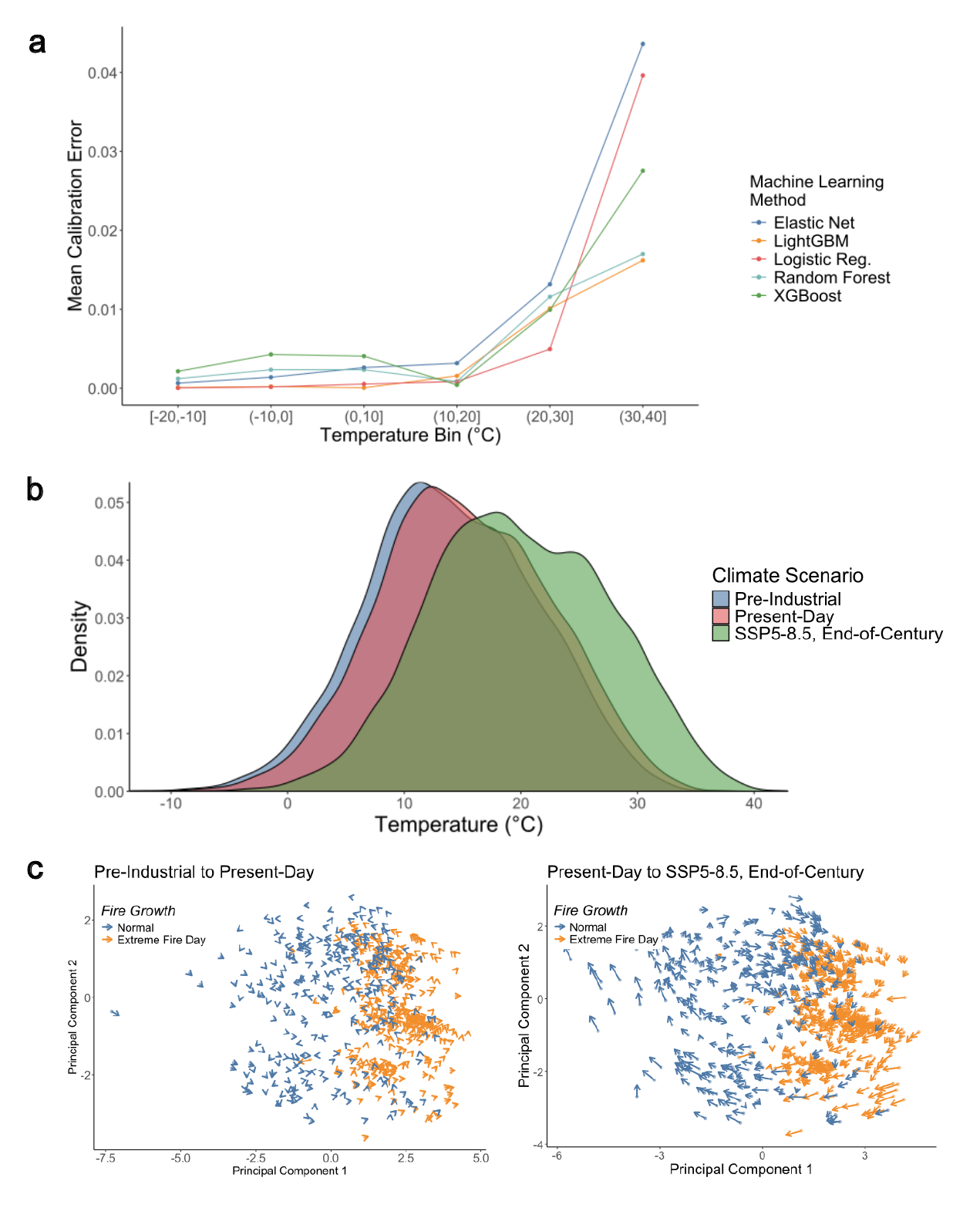}
    \caption[Predictive performance degrades when transporting to warmer temperatures.]{\textbf{Predictive performance degrades when transporting to warmer temperatures.} \textbf{a},  Mean calibration error when cross-validated on historically observed data, subgrouped by bins of temperature. \textbf{b}, Density plots illustrating temperature distribution shifts across climate scenarios. \textbf{c}, Principal components analysis of historically observed data, with arrows indicating direction and magnitude of covariate shift when compared to alternative climate scenarios.}
    \label{fig:fig5}
\end{figure}

\newpage
\begin{figure}[ht]
    \centering
    \includegraphics[width=1\linewidth]{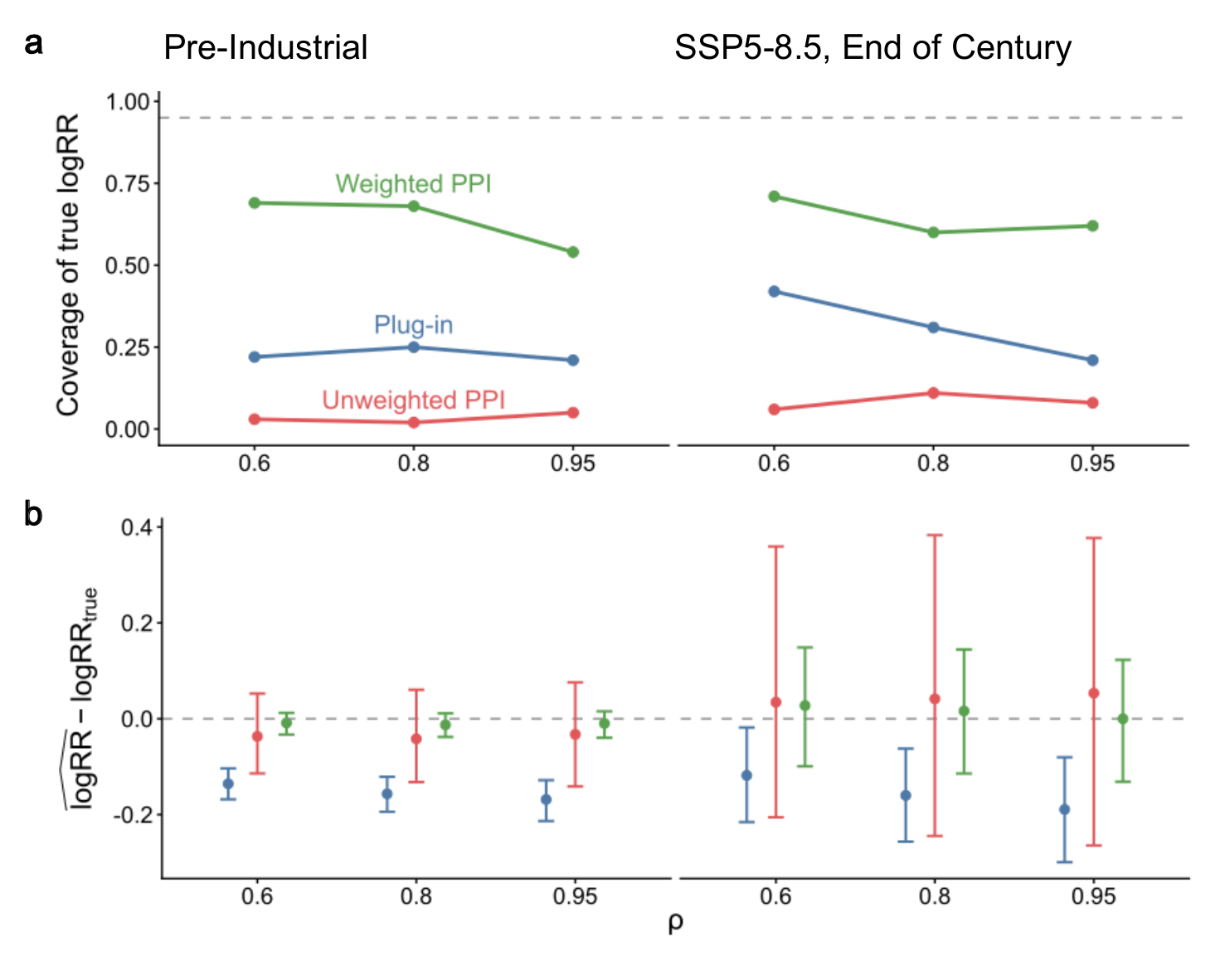}
    \caption[Absolute value of log risk ratio error, by model.]{\textbf{Absolute value of log risk ratio error, attribution estimation method.} \textbf{a}, Coverage of attribution estimation method over 3000 simulations per climate scenario. \textbf{b}, Median and IQR of the log risk ratio error estimates from 1000 simulations for each of the $\rho$ values, by attribution estimation method.}
    \label{fig:fig3}
\end{figure}

\newpage
\section{Tables}
\begin{table}[ht]
    \centering
    \caption[Aggregate fraction of attributable risk estimates (95\% Confidence Interval)]{Aggregate fraction of attributable risk estimates, weighted PPI method (95\% Confidence Interval)}
    \begin{tabular}{lcc}
    \toprule
    \textbf{Machine Learning} & \textbf{Pre-Industrial vs.} & \textbf{Present-Day vs. } \\
    \textbf{Method} & \textbf{Present-Day} & \textbf{SSP5-8.5, end-of-century} \\
    \midrule
        LightGBM & 0.154 (0.118, 0.200) & 0.442 (0.321, 0.521) \\
        Random Forest & 0.174 (0.139, 0.216) & 0.546 (0.468, 0.606) \\
        XGBoost & 0.212 (0.059, 0.266) & 0.450 (0.302, 0.526) \\
        Logistic Regression & 0.195 (0.156, 0.232) & 0.454 (0.360, 0.535) \\
        Elastic Net Regression & 0.196 (0.158, 0.236) & 0.474 (0.357, 0.555) \\
        Ensemble (Superlearner) & 0.175 (0.097, 0.270) & 0.523 (0.324, 0.591) \\
        \bottomrule
    \end{tabular}
    \label{tab:tab1}
\end{table}

\clearpage
\addcontentsline{toc}{section}{References}
\printbibliography

@article{breiman_random_2001,
	title = {Random {Forests}},
	volume = {45},
	issn = {1573-0565},
	url = {https://doi.org/10.1023/A:1010933404324},
	doi = {10.1023/A:1010933404324},
	language = {en},
	number = {1},
	urldate = {2025-11-24},
	journal = {Machine Learning},
	author = {Breiman, Leo},
	month = oct,
	year = {2001},
	pages = {5--32},
}

@article{angelopoulos_prediction-powered_2023,
	title = {Prediction-powered inference},
	volume = {382},
	url = {https://www.science.org/doi/full/10.1126/science.adi6000},
	doi = {10.1126/science.adi6000},
	number = {6671},
	urldate = {2025-11-24},
	journal = {Science},
	publisher = {American Association for the Advancement of Science},
	author = {Angelopoulos, Anastasios N. and Bates, Stephen and Fannjiang, Clara and Jordan, Michael I. and Zrnic, Tijana},
	month = nov,
	year = {2023},
	pages = {669--674},
}

@article{zou_regularization_2005,
	title = {Regularization and {Variable} {Selection} {Via} the {Elastic} {Net}},
	volume = {67},
	copyright = {https://academic.oup.com/journals/pages/open\_access/funder\_policies/chorus/standard\_publication\_model},
	issn = {1369-7412, 1467-9868},
	url = {https://academic.oup.com/jrsssb/article/67/2/301/7109482},
	doi = {10.1111/j.1467-9868.2005.00503.x},
	language = {en},
	number = {2},
	urldate = {2025-11-24},
	journal = {Journal of the Royal Statistical Society Series B: Statistical Methodology},
	author = {Zou, Hui and Hastie, Trevor},
	month = apr,
	year = {2005},
	pages = {301--320},
}

@inproceedings{chen_xgboost_2016,
	address = {San Francisco California USA},
	title = {{XGBoost}: {A} {Scalable} {Tree} {Boosting} {System}},
	isbn = {978-1-4503-4232-2},
	shorttitle = {{XGBoost}},
	url = {https://dl.acm.org/doi/10.1145/2939672.2939785},
	doi = {10.1145/2939672.2939785},
	language = {en},
	urldate = {2025-11-24},
	booktitle = {Proceedings of the 22nd {ACM} {SIGKDD} {International} {Conference} on {Knowledge} {Discovery} and {Data} {Mining}},
	publisher = {ACM},
	author = {Chen, Tianqi and Guestrin, Carlos},
	month = aug,
	year = {2016},
	pages = {785--794},
}

@inproceedings{ke_lightgbm_2017,
	title = {{LightGBM}: {A} {Highly} {Efficient} {Gradient} {Boosting} {Decision} {Tree}},
	volume = {30},
	shorttitle = {{LightGBM}},
	url = {https://proceedings.neurips.cc/paper/2017/hash/6449f44a102fde848669bdd9eb6b76fa-Abstract.html},
	urldate = {2025-11-24},
	booktitle = {Advances in {Neural} {Information} {Processing} {Systems}},
	publisher = {Curran Associates, Inc.},
	author = {Ke, Guolin and Meng, Qi and Finley, Thomas and Wang, Taifeng and Chen, Wei and Ma, Weidong and Ye, Qiwei and Liu, Tie-Yan},
	year = {2017},
}

@article{stott_attribution_2016,
	title = {Attribution of extreme weather and climate-related events},
	volume = {7},
	copyright = {© 2015 The Authors. WIREs Climate Change published by Wiley Periodicals, Inc.},
	issn = {1757-7799},
	url = {https://onlinelibrary.wiley.com/doi/abs/10.1002/wcc.380},
	doi = {10.1002/wcc.380},
	language = {en},
	number = {1},
	urldate = {2025-11-16},
	journal = {WIREs Climate Change},
	author = {Stott, Peter A. and Christidis, Nikolaos and Otto, Friederike E. L. and Sun, Ying and Vanderlinden, Jean-Paul and van Oldenborgh, Geert Jan and Vautard, Robert and von Storch, Hans and Walton, Peter and Yiou, Pascal and Zwiers, Francis W.},
	year = {2016},
	note = {\_eprint: https://wires.onlinelibrary.wiley.com/doi/pdf/10.1002/wcc.380},
	pages = {23--41},
}

@article{thompson_ethical_2015,
	title = {Ethical and normative implications of weather event attribution for policy discussions concerning loss and damage},
	volume = {133},
	issn = {1573-1480},
	url = {https://doi.org/10.1007/s10584-015-1433-z},
	doi = {10.1007/s10584-015-1433-z},
	language = {en},
	number = {3},
	urldate = {2025-11-16},
	journal = {Climatic Change},
	author = {Thompson, Allen and Otto, Friederike E. L.},
	month = dec,
	year = {2015},
	pages = {439--451},
}

@inproceedings{buolamwini_gender_2018,
	title = {Gender {Shades}: {Intersectional} {Accuracy} {Disparities} in {Commercial} {Gender} {Classification}},
	issn = {2640-3498},
	shorttitle = {Gender {Shades}},
	url = {https://proceedings.mlr.press/v81/buolamwini18a.html},
	language = {en},
	urldate = {2025-11-16},
	booktitle = {Proceedings of the 1st {Conference} on {Fairness}, {Accountability} and {Transparency}},
	publisher = {PMLR},
	author = {Buolamwini, Joy and Gebru, Timnit},
	month = jan,
	year = {2018},
	pages = {77--91},
}

@article{wong_external_2021,
	title = {External {Validation} of a {Widely} {Implemented} {Proprietary} {Sepsis} {Prediction} {Model} in {Hospitalized} {Patients}},
	volume = {181},
	issn = {2168-6106},
	url = {https://doi.org/10.1001/jamainternmed.2021.2626},
	doi = {10.1001/jamainternmed.2021.2626},
	number = {8},
	urldate = {2025-11-16},
	journal = {JAMA Internal Medicine},
	author = {Wong, Andrew and Otles, Erkin and Donnelly, John P. and Krumm, Andrew and McCullough, Jeffrey and DeTroyer-Cooley, Olivia and Pestrue, Justin and Phillips, Marie and Konye, Judy and Penoza, Carleen and Ghous, Muhammad and Singh, Karandeep},
	month = aug,
	year = {2021},
	pages = {1065--1070},
}

@article{obermeyer_dissecting_2019,
	title = {Dissecting racial bias in an algorithm used to manage the health of populations},
	volume = {366},
	url = {https://www.science.org/doi/10.1126/science.aax2342},
	doi = {10.1126/science.aax2342},
	number = {6464},
	urldate = {2025-11-16},
	journal = {Science},
	publisher = {American Association for the Advancement of Science},
	author = {Obermeyer, Ziad and Powers, Brian and Vogeli, Christine and Mullainathan, Sendhil},
	month = oct,
	year = {2019},
	pages = {447--453},
}

@inproceedings{corbett-davies_algorithmic_2017,
	address = {New York, NY, USA},
	series = {{KDD} '17},
	title = {Algorithmic {Decision} {Making} and the {Cost} of {Fairness}},
	isbn = {978-1-4503-4887-4},
	url = {https://doi.org/10.1145/3097983.3098095},
	doi = {10.1145/3097983.3098095},
	urldate = {2025-11-15},
	booktitle = {Proceedings of the 23rd {ACM} {SIGKDD} {International} {Conference} on {Knowledge} {Discovery} and {Data} {Mining}},
	publisher = {Association for Computing Machinery},
	author = {Corbett-Davies, Sam and Pierson, Emma and Feller, Avi and Goel, Sharad and Huq, Aziz},
	month = aug,
	year = {2017},
	pages = {797--806},
}

@article{dressel_accuracy_2018,
	title = {The accuracy, fairness, and limits of predicting recidivism},
	volume = {4},
	url = {https://www.science.org/doi/10.1126/sciadv.aao5580},
	doi = {10.1126/sciadv.aao5580},
	number = {1},
	urldate = {2025-11-16},
	journal = {Science Advances},
	publisher = {American Association for the Advancement of Science},
	author = {Dressel, Julia and Farid, Hany},
	month = jan,
	year = {2018},
	pages = {eaao5580},
}

@misc{jimenez-esteve_ai-driven_2024,
	title = {{AI}-driven weather forecasts enable anticipated attribution of extreme events to human-made climate change},
	url = {http://arxiv.org/abs/2408.16433},
	doi = {10.48550/arXiv.2408.16433},
	urldate = {2025-11-12},
	publisher = {arXiv},
	author = {Jiménez-Esteve, Bernat and Barriopedro, David and Johnson, Juan Emmanuel and Garcia-Herrera, Ricardo},
	month = aug,
	year = {2024},
	note = {arXiv:2408.16433 [physics]},
}

@article{loridan_reask_2025,
	title = {Reask {UTC}: a machine learning modeling framework to generate climate-connected tropical cyclone event sets globally},
	volume = {25},
	issn = {1561-8633},
	shorttitle = {Reask {UTC}},
	url = {https://nhess.copernicus.org/articles/25/2863/2025/},
	doi = {10.5194/nhess-25-2863-2025},
	language = {English},
	number = {8},
	urldate = {2025-11-12},
	journal = {Natural Hazards and Earth System Sciences},
	publisher = {Copernicus GmbH},
	author = {Loridan, Thomas and Bruneau, Nicolas},
	month = aug,
	year = {2025},
	pages = {2863--2884},
}

@article{perkins-kirkpatrick_attribution_2022,
	title = {On the attribution of the impacts of extreme weather events to anthropogenic climate change},
	volume = {17},
	issn = {1748-9326},
	url = {https://doi.org/10.1088/1748-9326/ac44c8},
	doi = {10.1088/1748-9326/ac44c8},
	language = {en},
	number = {2},
	urldate = {2025-11-12},
	journal = {Environmental Research Letters},
	publisher = {IOP Publishing},
	author = {Perkins-Kirkpatrick, S E and Stone, D A and Mitchell, D M and Rosier, S and King, A D and Lo, Y T E and Pastor-Paz, J and Frame, D and Wehner, M},
	month = jan,
	year = {2022},
	pages = {024009},
}

@article{otto_attribution_2017,
	title = {Attribution of {Weather} and {Climate} {Events}},
	volume = {42},
	issn = {1543-5938, 1545-2050},
	url = {https://www.annualreviews.org/content/journals/10.1146/annurev-environ-102016-060847},
	doi = {10.1146/annurev-environ-102016-060847},
	language = {en},
	number = {Volume 42, 2017},
	urldate = {2025-11-12},
	journal = {Annual Review of Environment and Resources},
	publisher = {Annual Reviews},
	author = {Otto, Friederike E. L.},
	month = oct,
	year = {2017},
	pages = {627--646},
}

@article{lloyd_climate_2021,
	title = {Climate change attribution and legal contexts: evidence and the role of storylines},
	volume = {167},
	issn = {1573-1480},
	shorttitle = {Climate change attribution and legal contexts},
	url = {https://doi.org/10.1007/s10584-021-03177-y},
	doi = {10.1007/s10584-021-03177-y},
	language = {en},
	number = {3},
	urldate = {2025-11-12},
	journal = {Climatic Change},
	author = {Lloyd, Elisabeth A. and Shepherd, Theodore G.},
	month = aug,
	year = {2021},
	pages = {28},
}

@article{adam_climate_2011,
	title = {Climate change in court},
	volume = {1},
	copyright = {2011 Springer Nature Limited},
	issn = {1758-6798},
	url = {https://www.nature.com/articles/nclimate1131},
	doi = {10.1038/nclimate1131},
	language = {en},
	number = {3},
	urldate = {2025-11-12},
	journal = {Nature Climate Change},
	publisher = {Nature Publishing Group},
	author = {Adam, David},
	month = jun,
	year = {2011},
	pages = {127--130},
}

@article{clarke_inventories_2021,
	title = {Inventories of extreme weather events and impacts: {Implications} for loss and damage from and adaptation to climate extremes},
	volume = {32},
	issn = {2212-0963},
	shorttitle = {Inventories of extreme weather events and impacts},
	url = {https://www.sciencedirect.com/science/article/pii/S2212096321000140},
	doi = {10.1016/j.crm.2021.100285},
	urldate = {2025-11-12},
	journal = {Climate Risk Management},
	author = {Clarke, Ben J. and E. L. Otto, Friederike and Jones, Richard G.},
	month = jan,
	year = {2021},
	pages = {100285},
}

@article{marjanac_extreme_2018,
	title = {Extreme weather event attribution science and climate change litigation: an essential step in the causal chain?},
	volume = {36},
	issn = {0264-6811},
	shorttitle = {Extreme weather event attribution science and climate change litigation},
	url = {https://doi.org/10.1080/02646811.2018.1451020},
	doi = {10.1080/02646811.2018.1451020},
	number = {3},
	urldate = {2025-11-12},
	journal = {Journal of Energy \& Natural Resources Law},
	publisher = {Routledge},
	author = {Marjanac, Sophie and Patton, Lindene},
	month = jul,
	year = {2018},
	note = {\_eprint: https://doi.org/10.1080/02646811.2018.1451020},
	pages = {265--298},
}

@article{burger_law_2020,
	title = {The {Law} and {Science} of {Climate} {Change} {Attribution}},
	volume = {45},
	copyright = {Copyright (c) 2020 Michael Burger, Jessica Wentz, Radley Horton},
	issn = {2837-5297},
	url = {https://journals.library.columbia.edu/index.php/cjel/article/view/4730},
	doi = {10.7916/cjel.v45i1.4730},
	language = {en},
	number = {1},
	urldate = {2025-11-12},
	journal = {Columbia Journal of Environmental Law},
	author = {Burger, Michael and Wentz, Jessica and Horton, Radley},
	month = feb,
	year = {2020},
}

@article{lam_learning_2023,
	title = {Learning skillful medium-range global weather forecasting},
	volume = {382},
	url = {https://www.science.org/doi/10.1126/science.adi2336},
	doi = {10.1126/science.adi2336},
	number = {6677},
	urldate = {2025-11-12},
	journal = {Science},
	publisher = {American Association for the Advancement of Science},
	author = {Lam, Remi and Sanchez-Gonzalez, Alvaro and Willson, Matthew and Wirnsberger, Peter and Fortunato, Meire and Alet, Ferran and Ravuri, Suman and Ewalds, Timo and Eaton-Rosen, Zach and Hu, Weihua and Merose, Alexander and Hoyer, Stephan and Holland, George and Vinyals, Oriol and Stott, Jacklynn and Pritzel, Alexander and Mohamed, Shakir and Battaglia, Peter},
	month = dec,
	year = {2023},
	pages = {1416--1421},
}

@misc{vaughan_aardvark_2024,
	title = {Aardvark {Weather}: end-to-end data-driven weather forecasting},
	shorttitle = {Aardvark {Weather}},
	url = {http://arxiv.org/abs/2404.00411},
	doi = {10.48550/arXiv.2404.00411},
	urldate = {2025-11-12},
	publisher = {arXiv},
	author = {Vaughan, Anna and Markou, Stratis and Tebbutt, Will and Requeima, James and Bruinsma, Wessel P. and Andersson, Tom R. and Herzog, Michael and Lane, Nicholas D. and Hosking, J. Scott and Turner, Richard E.},
	month = apr,
	year = {2024},
	note = {arXiv:2404.00411 [physics]
version: 1},
}

@article{price_probabilistic_2025,
	title = {Probabilistic weather forecasting with machine learning},
	volume = {637},
	copyright = {2024 The Author(s)},
	issn = {1476-4687},
	url = {https://www.nature.com/articles/s41586-024-08252-9},
	doi = {10.1038/s41586-024-08252-9},
	language = {en},
	number = {8044},
	urldate = {2025-11-12},
	journal = {Nature},
	publisher = {Nature Publishing Group},
	author = {Price, Ilan and Sanchez-Gonzalez, Alvaro and Alet, Ferran and Andersson, Tom R. and El-Kadi, Andrew and Masters, Dominic and Ewalds, Timo and Stott, Jacklynn and Mohamed, Shakir and Battaglia, Peter and Lam, Remi and Willson, Matthew},
	month = jan,
	year = {2025},
	pages = {84--90},
}

@article{charlton-perez_ai_2024,
	title = {Do {AI} models produce better weather forecasts than physics-based models? {A} quantitative evaluation case study of {Storm} {Ciarán}},
	volume = {7},
	copyright = {2024 The Author(s)},
	issn = {2397-3722},
	shorttitle = {Do {AI} models produce better weather forecasts than physics-based models?},
	url = {https://www.nature.com/articles/s41612-024-00638-w},
	doi = {10.1038/s41612-024-00638-w},
	language = {en},
	number = {1},
	urldate = {2025-11-12},
	journal = {npj Climate and Atmospheric Science},
	publisher = {Nature Publishing Group},
	author = {Charlton-Perez, Andrew J. and Dacre, Helen F. and Driscoll, Simon and Gray, Suzanne L. and Harvey, Ben and Harvey, Natalie J. and Hunt, Kieran M. R. and Lee, Robert W. and Swaminathan, Ranjini and Vandaele, Remy and Volonté, Ambrogio},
	month = apr,
	year = {2024},
	pages = {93},
}

@article{kurth_fourcastnet_nodate,
	title = {{FourCastNet}: {Accelerating} {Global} {High}-{Resolution} {Weather} {Forecasting} using {Adaptive} {Fourier} {Neural} {Operators}},
	shorttitle = {{FourCastNet}},
	url = {https://authors.library.caltech.edu/records/98c3k-er634},
	doi = {10.48550/arXiv.2208.05419},
	language = {en},
	urldate = {2025-11-12},
	author = {Kurth, Thorsten and Subramanian, Shashank and Harrington, Peter and Pathak, Jaideep and Mardani, Morteza and Hall, David and Miele, Andrea and Kashinath, Karthik and Anandkumar, Animashree},
}

@article{liu_evaluation_2024,
	title = {Evaluation of five global {AI} models for predicting weather in {Eastern} {Asia} and {Western} {Pacific}},
	volume = {7},
	copyright = {2024 The Author(s)},
	issn = {2397-3722},
	url = {https://www.nature.com/articles/s41612-024-00769-0},
	doi = {10.1038/s41612-024-00769-0},
	language = {en},
	number = {1},
	urldate = {2025-11-12},
	journal = {npj Climate and Atmospheric Science},
	publisher = {Nature Publishing Group},
	author = {Liu, Cheng-Chin and Hsu, Kathryn and Peng, Melinda S. and Chen, Der-Song and Chang, Pao-Liang and Hsiao, Ling-Feng and Fong, Chin-Tzu and Hong, Jing-Shan and Cheng, Chia-Ping and Lu, Kuo-Chen and Chen, Chia-Rong and Kuo, Hung-Chi},
	month = sep,
	year = {2024},
	pages = {221},
}

@article{bi_accurate_2023,
	title = {Accurate medium-range global weather forecasting with {3D} neural networks},
	volume = {619},
	copyright = {2023 The Author(s)},
	issn = {1476-4687},
	url = {https://www.nature.com/articles/s41586-023-06185-3},
	doi = {10.1038/s41586-023-06185-3},
	language = {en},
	number = {7970},
	urldate = {2025-11-12},
	journal = {Nature},
	publisher = {Nature Publishing Group},
	author = {Bi, Kaifeng and Xie, Lingxi and Zhang, Hengheng and Chen, Xin and Gu, Xiaotao and Tian, Qi},
	month = jul,
	year = {2023},
	pages = {533--538},
}

@article{camps-valls_artificial_2025,
	title = {Artificial intelligence for modeling and understanding extreme weather and climate events},
	volume = {16},
	copyright = {2025 The Author(s)},
	issn = {2041-1723},
	url = {https://www.nature.com/articles/s41467-025-56573-8},
	doi = {10.1038/s41467-025-56573-8},
	language = {en},
	number = {1},
	urldate = {2025-11-12},
	journal = {Nature Communications},
	publisher = {Nature Publishing Group},
	author = {Camps-Valls, Gustau and Fernández-Torres, Miguel-Ángel and Cohrs, Kai-Hendrik and Höhl, Adrian and Castelletti, Andrea and Pacal, Aytac and Robin, Claire and Martinuzzi, Francesco and Papoutsis, Ioannis and Prapas, Ioannis and Pérez-Aracil, Jorge and Weigel, Katja and Gonzalez-Calabuig, Maria and Reichstein, Markus and Rabel, Martin and Giuliani, Matteo and Mahecha, Miguel D. and Popescu, Oana-Iuliana and Pellicer-Valero, Oscar J. and Ouala, Said and Salcedo-Sanz, Sancho and Sippel, Sebastian and Kondylatos, Spyros and Happé, Tamara and Williams, Tristan},
	month = feb,
	year = {2025},
	pages = {1919},
}

@article{lott_evaluating_2016,
	chapter = {Journal of Climate},
	title = {Evaluating {Simulated} {Fraction} of {Attributable} {Risk} {Using} {Climate} {Observations}},
	url = {https://journals.ametsoc.org/view/journals/clim/29/12/jcli-d-15-0566.1.xml},
	doi = {10.1175/JCLI-D-15-0566.1},
	language = {en},
	urldate = {2025-11-12},
	author = {Lott, Fraser C. and Stott, Peter A.},
	month = jun,
	year = {2016},
}

@misc{dai_be_2025,
	title = {Be {Intentional} {About} {Fairness}!: {Fairness}, {Size}, and {Multiplicity} in the {Rashomon} {Set}},
	shorttitle = {Be {Intentional} {About} {Fairness}!},
	url = {http://arxiv.org/abs/2501.15634},
	doi = {10.48550/arXiv.2501.15634},
	urldate = {2025-11-07},
	publisher = {arXiv},
	author = {Dai, Gordon and Ravishankar, Pavan and Yuan, Rachel and Neill, Daniel B. and Black, Emily},
	month = jan,
	year = {2025},
	note = {arXiv:2501.15634 [cs]},
}

@article{brown_climate_2023,
	title = {Climate warming increases extreme daily wildfire growth risk in {California}},
	volume = {621},
	copyright = {2023 The Author(s), under exclusive licence to Springer Nature Limited},
	issn = {1476-4687},
	url = {https://www.nature.com/articles/s41586-023-06444-3},
	doi = {10.1038/s41586-023-06444-3},
	language = {en},
	number = {7980},
	urldate = {2025-11-06},
	journal = {Nature},
	publisher = {Nature Publishing Group},
	author = {Brown, Patrick T. and Hanley, Holt and Mahesh, Ankur and Reed, Colorado and Strenfel, Scott J. and Davis, Steven J. and Kochanski, Adam K. and Clements, Craig B.},
	month = sep,
	year = {2023},
	pages = {760--766},
}

@article{trok_machine_2024,
	title = {Machine learning–based extreme event attribution},
	volume = {10},
	url = {https://www.science.org/doi/10.1126/sciadv.adl3242},
	doi = {10.1126/sciadv.adl3242},
	number = {34},
	urldate = {2025-11-06},
	journal = {Science Advances},
	publisher = {American Association for the Advancement of Science},
	author = {Trok, Jared T. and Barnes, Elizabeth A. and Davenport, Frances V. and Diffenbaugh, Noah S.},
	month = aug,
	year = {2024},
	pages = {eadl3242},
}

@misc{ovadia_can_2019,
	title = {Can {You} {Trust} {Your} {Model}'s {Uncertainty}? {Evaluating} {Predictive} {Uncertainty} {Under} {Dataset} {Shift}},
	shorttitle = {Can {You} {Trust} {Your} {Model}'s {Uncertainty}?},
	url = {http://arxiv.org/abs/1906.02530},
	doi = {10.48550/arXiv.1906.02530},
	urldate = {2025-11-06},
	publisher = {arXiv},
	author = {Ovadia, Yaniv and Fertig, Emily and Ren, Jie and Nado, Zachary and Sculley, D. and Nowozin, Sebastian and Dillon, Joshua V. and Lakshminarayanan, Balaji and Snoek, Jasper},
	month = dec,
	year = {2019},
	note = {arXiv:1906.02530 [stat]},
}

@inproceedings{roth_reconciling_2023,
	address = {New York, NY, USA},
	series = {{FAccT} '23},
	title = {Reconciling {Individual} {Probability} {Forecasts}},
	isbn = {979-8-4007-0192-4},
	url = {https://dl.acm.org/doi/10.1145/3593013.3593980},
	doi = {10.1145/3593013.3593980},
	urldate = {2025-11-06},
	booktitle = {Proceedings of the 2023 {ACM} {Conference} on {Fairness}, {Accountability}, and {Transparency}},
	publisher = {Association for Computing Machinery},
	author = {Roth, Aaron and Tolbert, Alexander and Weinstein, Scott},
	month = jun,
	year = {2023},
	pages = {101--110},
}

@misc{behzad_reconciling_2025,
	title = {Reconciling {Predictive} {Multiplicity} in {Practice}},
	url = {http://arxiv.org/abs/2501.16549},
	doi = {10.48550/arXiv.2501.16549},
	urldate = {2025-11-06},
	publisher = {arXiv},
	author = {Behzad, Tina and Casacuberta, Sílvia and Diana, Emily Ruth and Tolbert, Alexander Williams},
	month = jan,
	year = {2025},
	note = {arXiv:2501.16549 [cs]},
}

@article{naveau_statistical_2020,
	title = {Statistical {Methods} for {Extreme} {Event} {Attribution} in {Climate} {Science}},
	volume = {7},
	issn = {2326-8298, 2326-831X},
	url = {https://www.annualreviews.org/content/journals/10.1146/annurev-statistics-031219-041314},
	doi = {10.1146/annurev-statistics-031219-041314},
	language = {en},
	number = {Volume 7, 2020},
	urldate = {2025-11-06},
	journal = {Annual Review of Statistics and Its Application},
	publisher = {Annual Reviews},
	author = {Naveau, Philippe and Hannart, Alexis and Ribes, Aurélien},
	month = mar,
	year = {2020},
	pages = {89--110},
}

@article{dorie_automated_2019,
	title = {Automated versus {Do}-{It}-{Yourself} {Methods} for {Causal} {Inference}: {Lessons} {Learned} from a {Data} {Analysis} {Competition}},
	volume = {34},
	issn = {0883-4237},
	shorttitle = {Automated versus {Do}-{It}-{Yourself} {Methods} for {Causal} {Inference}},
	url = {https://projecteuclid.org/journals/statistical-science/volume-34/issue-1/Automated-versus-Do-It-Yourself-Methods-for-Causal-Inference/10.1214/18-STS667.full},
	doi = {10.1214/18-STS667},
	language = {en},
	number = {1},
	urldate = {2026-07-20},
	journal = {Statistical Science},
	author = {Dorie, Vincent and Hill, Jennifer and Shalit, Uri and Scott, Marc and Cervone, Dan},
	month = feb,
	year = {2019},
}

@article{polley_super_2010,
	title = {Super {Learner} {In} {Prediction}},
	url = {https://biostats.bepress.com/ucbbiostat/paper266},
	journal = {U.C. Berkeley Division of Biostatistics Working Paper Series},
	author = {Polley, Eric and Laan, Mark van der},
	month = may,
	year = {2010},
}

@article{laan_super_nodate,
	title = {Super {Learner}},
	volume = {6},
	copyright = {De Gruyter expressly reserves the right to use all content for commercial text and data mining within the meaning of Section 44b of the German Copyright Act.},
	issn = {1544-6115},
	url = {https://www.degruyterbrill.com/document/doi/10.2202/1544-6115.1309/html?lang=en},
	doi = {10.2202/1544-6115.1309},
	language = {en},
	number = {1},
	urldate = {2026-05-18},
	journal = {Statistical Applications in Genetics and Molecular Biology},
	publisher = {De Gruyter},
	author = {Laan, Mark J. van der and Polley, Eric C. and Hubbard, Alan E.},
}

@misc{noauthor_chapter_nodate,
	title = {Chapter 11: {Weather} and {Climate} {Extreme} {Events} in a {Changing} {Climate}},
	shorttitle = {Chapter 11},
	url = {https://www.ipcc.ch/report/ar6/wg1/chapter/chapter-11/},
	language = {en},
	urldate = {2026-04-20},
}

@misc{lucena_spline-based_2018,
	title = {Spline-{Based} {Probability} {Calibration}},
	url = {http://arxiv.org/abs/1809.07751},
	doi = {10.48550/arXiv.1809.07751},
	urldate = {2026-04-20},
	publisher = {arXiv},
	author = {Lucena, Brian},
	month = sep,
	year = {2018},
	note = {arXiv:1809.07751 [stat]},
}

@article{niculescu-mizil_obtaining_nodate,
	title = {Obtaining {Calibrated} {Probabilities} from {Boosting}},
	language = {en},
	author = {Niculescu-Mizil, Alexandru and Caruana, Rich},
}

@inproceedings{tomani_post-hoc_2021,
	address = {Nashville, TN, USA},
	title = {Post-hoc {Uncertainty} {Calibration} for {Domain} {Drift} {Scenarios}},
	copyright = {https://ieeexplore.ieee.org/Xplorehelp/downloads/license-information/IEEE.html},
	isbn = {978-1-6654-4509-2},
	url = {https://ieeexplore.ieee.org/document/9577789/},
	doi = {10.1109/CVPR46437.2021.00999},
	language = {en},
	urldate = {2026-04-20},
	booktitle = {2021 {IEEE}/{CVF} {Conference} on {Computer} {Vision} and {Pattern} {Recognition} ({CVPR})},
	publisher = {IEEE},
	author = {Tomani, Christian and Gruber, Sebastian and Erdem, Muhammed Ebrar and Cremers, Daniel and Buettner, Florian},
	month = jun,
	year = {2021},
	pages = {10119--10127},
}

@inproceedings{ma_meta-cal_2021,
	title = {Meta-{Cal}: {Well}-controlled {Post}-hoc {Calibration} by {Ranking}},
	issn = {2640-3498},
	shorttitle = {Meta-{Cal}},
	url = {https://proceedings.mlr.press/v139/ma21a.html},
	language = {en},
	urldate = {2026-04-20},
	booktitle = {Proceedings of the 38th {International} {Conference} on {Machine} {Learning}},
	publisher = {PMLR},
	author = {Ma, Xingchen and Blaschko, Matthew B.},
	month = jul,
	year = {2021},
	pages = {7235--7245},
}

@article{platt_probabilistic_2000,
	title = {Probabilistic {Outputs} for {Support} {Vector} {Machines} and {Comparisons} to {Regularized} {Likelihood} {Methods}},
	volume = {10},
	journal = {Adv. Large Margin Classif.},
	author = {Platt, John},
	month = jun,
	year = {2000},
}

@article{collins_evaluation_2024,
	title = {Evaluation of clinical prediction models (part 1): from development to external validation},
	volume = {384},
	issn = {0959-8138},
	shorttitle = {Evaluation of clinical prediction models (part 1)},
	url = {https://pmc.ncbi.nlm.nih.gov/articles/PMC10772854/},
	doi = {10.1136/bmj-2023-074819},
	urldate = {2026-04-20},
	journal = {The BMJ},
	author = {Collins, Gary S and Dhiman, Paula and Ma, Jie and Schlussel, Michael M and Archer, Lucinda and Van Calster, Ben and Harrell, Frank E and Martin, Glen P and Moons, Karel G M and van Smeden, Maarten and Sperrin, Matthew and Bullock, Garrett S and Riley, Richard D},
	month = jan,
	year = {2024},
	pages = {e074819},
}

@article{callahan_increasing_2026,
	title = {Increasing risk of mass human heat mortality if historical weather patterns recur},
	volume = {16},
	url = {https://ui.adsabs.harvard.edu/abs/2026NatCC..16...26C},
	doi = {10.1038/s41558-025-02480-1},
	urldate = {2026-04-19},
	journal = {Nature Climate Change},
	publisher = {Springer},
	author = {Callahan, Christopher W. and Trok, Jared and Wilson, Andrew J. and Gould, Carlos F. and Heft-Neal, Sam and Diffenbaugh, Noah S. and Burke, Marshall},
	month = jan,
	year = {2026},
	note = {ADS Bibcode: 2026NatCC..16...26C},
	pages = {26--32},
}

@inproceedings{ben-david_analysis_2006,
	title = {Analysis of {Representations} for {Domain} {Adaptation}},
	volume = {19},
	url = {https://papers.nips.cc/paper_files/paper/2006/hash/b1b0432ceafb0ce714426e9114852ac7-Abstract.html},
	urldate = {2026-04-19},
	booktitle = {Advances in {Neural} {Information} {Processing} {Systems}},
	publisher = {MIT Press},
	author = {Ben-David, Shai and Blitzer, John and Crammer, Koby and Pereira, Fernando},
	year = {2006},
}

@book{jr_applied_2013,
	title = {Applied {Logistic} {Regression}},
	isbn = {978-1-118-54835-6},
	language = {en},
	publisher = {John Wiley \& Sons},
	author = {Jr, David W. Hosmer and Lemeshow, Stanley and Sturdivant, Rodney X.},
	month = feb,
	year = {2013},
	note = {Google-Books-ID: bRoxQBIZRd4C},
}

@inproceedings{rahimi_random_2007,
	title = {Random {Features} for {Large}-{Scale} {Kernel} {Machines}},
	volume = {20},
	url = {https://papers.nips.cc/paper_files/paper/2007/hash/013a006f03dbc5392effeb8f18fda755-Abstract.html},
	urldate = {2026-08-11},
	booktitle = {Advances in {Neural} {Information} {Processing} {Systems}},
	publisher = {Curran Associates, Inc.},
	author = {Rahimi, Ali and Recht, Benjamin},
	year = {2007},
}

\newpage
\section{Materials and Methods}\label{methods}
\subsection{Data}
To replicate experiments from Brown et al. \autocite{brown_climate_2023}, we use their data on wildfire days across California from 2003-2020 (n = 17,910 fire days). Wildfire days may come from separate dates within the same wildfire event. Their data are obtained from a variety of publicly available sources. Wildfire days are sourced from MODIS satellite estimates from NASA. Predictor variables are from reanalysis produced from the National Center for Atmospheric Research’s Weather Research Forecasting model. Climate simulation data are from CMIP6.

Our variable of interest is extreme daily growth, specifically defined as wildfire growth over 10,000 acres in a single day. There are 380 extreme daily growth days in our observed dataset. Our predictor variables fall into two main categories: weather and land. Weather variables are time-varying and include temperature, vapor pressure deficit, 100 hour dead fuel moisture and 1000 hour dead fuel moisture, precipitation, and wind speed. Land variables are non-time varying and include land use category (forest, shrub, or savanna/grassland), land aspect (cardinal direction), land slope, elevation above sea level, and vegetation fraction. 

Across predictor datasets that represent different climate scenarios, the land, precipitation, and wind speed variables are held constant. As described in Brown et al.\autocite{brown_climate_2023}, temperature is changed using general circulation models to represent a pre-industrial time period and the SSP5-8.5 future emission scenario. They propagate their predicted temperature changes into the vapor pressure deficit and dead fuel moisture variables using physics-based calculations. As a result of this data formation, each wildfire day in our present-day dataset has a corresponding wildfire day in the pre-industrial and SSP5-8.5 future emission scenarios with matching land, precipitation, and wind speed values. 

\subsection{Methodology}
Our overall approach consists of training machine learning models to predict the probability of extreme daily growth of wildfires in California from 2003-2020. As in Brown et al.\autocite{brown_climate_2023}, summarized below, we then used predictions from these models under different climate scenarios to assess potential attribution to anthropogenic emissions. From there, we assessed three sources of uncertainty in our estimations: model multiplicity, machine learning error, and distribution shift in predictors under different climate scenarios (Figure \ref{fig:fig1}). 

\subsubsection{Predicting probability of extreme daily growth}
We train and tune models using cross validation on the full present-day dataset to predict the probability of extreme daily growth in the pre-industrial and SSP5-8.5 scenarios ($\hat{p^{(1)}}$). We train our models using five different machine learning algorithms: LightGBM \autocite{ke_lightgbm_2017}, Random Forest \autocite{breiman_random_2001}, XGBoost \autocite{chen_xgboost_2016}, Logistic Regression \autocite{jr_applied_2013}, and Elastic Net Regression \autocite{zou_regularization_2005}. We select these algorithms to represent a combination of machine learning models used in prior studies (e.g. random forest), more modern machine learning techniques (e.g. lightGBM and XGBoost), and more traditional statistical methods (e.g. logistic regression and elastic net regression). We additionally include an ensembled model using the SuperLearner algorithm \autocite{laan_super_nodate, polley_super_2010}, which combines all five previously mentioned machine learning models as base learners. In total, for each climate scenario, we have six sets of out-of-sample predictions for each counterfactual time period, each set representing a different machine learning algorithm. We evaluate the performance of our models in the present-day dataset using AUC, Brier score, Brier skill score, and mean calibration error, described in a subsequent section.

\subsubsection{Calibrating predictions post-hoc}
Post-hoc calibration adjusts the predicted probabilities from a machine learning model to better reflect the true outcome frequency.\autocite{collins_evaluation_2024,ma_meta-cal_2021,tomani_post-hoc_2021} More specifically, post-hoc calibration aims to match the observed outcome frequency and the predicted probability across points with the same predicted probability. Post-hoc calibration is commonly used to correct potential overconfidence, improve reliability, and correct bias in our machine learning models. 

To calibrate our predictions, we learn a calibration model using the cross-validated predictions from each machine learning method in the present-day dataset as our input and their observed outcomes as our output. This results in a separate calibration model for each machine learning method. We apply this learned calibration model to our predicted probabilities from the corresponding machine learning method in the counterfactual scenarios to get a set of calibrated predictions. We then repeated this process with four different calibration methods to represent methods that could reasonably be chosen in practice: logistic calibration (or Platt scaling)\autocite{platt_probabilistic_2000}, isotonic calibration\autocite{niculescu-mizil_obtaining_nodate}, natural spline logistic calibration, and I-Spline logistic calibration.\autocite{lucena_spline-based_2018}

\subsubsection{Estimating Fraction of Attributable Risk (FAR)}
Fraction of attributable risk is a metric used to assess potential attribution to anthropogenic emissions in climate science. It can be represented as:
\[ FAR =1-\frac{E[Y^{(0)}]}{E[Y^{(1)}]}\]

For calculating and reporting attribution error, we report in terms of risk ratio, a related metric to FAR:
\[ RR =\frac{E[Y^{(1)}]}{E[Y^{(0)}]}\]

In scenarios with small values of $E[Y^{(1)}]$, the FAR value may become infinitely large. Therefore, we also report the average treatment effect (ATE) in the supplement, which is proportional to FAR. Average treatment effect is calculated as:
\[ ATE = E[Y^{(1)}] - E[Y^{(0)}]\]

These quantities can be calculated for individual fire days using estimated probability of extreme daily growth on day $i$ in both the observed and counterfactual datasets. 
They can also be calculated as aggregate quantities that represent the average attribution of changes in extreme fire growth to anthropogenic emissions. We focus on these aggregate attribution estimates through most of this analysis. We use the observed proportion of extreme daily growth, 380/17910, and the mean of the predicted probabilities in the counterfactual scenario to calculate these quantities.

To quantify model sensitivity to design choices, we calculate the range factor for risk ratio estimates for each individual day and for each climate scenario. Range factor is calculated as the ratio of the maximum RR estimate to the minimum RR estimate across the six machine learning model estimates. It can be represented as:
\[Range \space Factor= \frac{\max{\hat{RR_i}}}{\min{\hat{RR_i}}}\]

\subsubsection{Estimating uncertainty}
We use a bootstrap method to estimate a 95\% confidence interval for our estimates of aggregate ATE and FAR. We sample 17,910 matching days with replacement from our present-day and counterfactual datasets, retrain our machine learning models, and estimate RR. We repeat this process 1,000 times, taking the 2.5 and 97.5 percentile values of our 1,000 bootstrapped estimates to get the 95\% confidence interval.

\subsubsection{Assessing predictive performance metrics' correlation with accuracy of aggregate FAR estimates via simulation}
To assess the correlation between predictive performance metrics and accuracy of our aggregate estimates, we generate datasets where we know the attribution effect and implement the same model training, calibration, and estimation process. We generate datasets by treating our predicted probabilities of extreme daily growth in the present-day and counterfactual scenarios as the true probabilities of extreme daily growth ($p(X)^{(0)}$ and $p(X)^{(1)}$). The probability of extreme daily growth used to calculate FAR, RR, and ATE is the mean of these predicted probabilities. We have six different “truth” scenarios representing each machine learning method of generating the predicted probabilities, and generate 300 datasets per scenario.

Specifically, for each day in our original dataset, we use our predicted probability of extreme daily growth in the present-day scenario to generate 300 Bernoulli values that represent outcomes in 300 present-day alternate realities. The aggregated Bernoulli values represent our new outcomes; the predictor variables remain the same as the present-day scenario across all 300 datasets. We train our machine learning models on the present-day datasets and estimate the average treatment effect and fraction of attributable risk in each simulated reality using the same methods as above. We calculate the residual difference from the true values to compare performance of the different machine learning methods and performance of the effect estimation methods. Formal representation described in the supplementary information.

\subsubsection{Calculating performance metrics and their correlation with accuracy of FAR}
To calculate performance metrics for our models, we used cross-validated predictions and the true outcomes from our present-day dataset. Specifically, we trained several machine learning models on our present-day dataset to predict the probability of extreme daily growth ($\hat{p(X)^{(0)}}$) using land and weather variables. We build separate models to make out-of-sample predictions for each three-year time period (2003-2005, 2006-2008, 2009-2011, 2012-2014, 2015-2017, 2018-2020). Each model is trained and tuned using cross-validation on the remaining five time periods (e.g. the model for prediction in 2003-2005 is trained on data from 2006-2020). We split our sample based on time period to avoid training and validating on days from the same fire. 

Using 500 of our simulated datasets -- 100 for each ``truth" scenario excluding the ensemble method -- we evaluated the correlation between standard predictive performance metrics and the accuracy of our aggregate FAR estimates to assess the best performance metrics for choosing a model in our context. We use absolute log risk ratio error as a measure of attribution accuracy, which can be represented as: $|\log{(RR_{error)}}| =|\log{\frac{\hat{Y^{(0)}}}{\hat{p(X)^{(1)}}}} - \log{\frac{Y^{(0)}}{p(X)^{(1)}}}|$. We calculated the area under the ROC curve (AUC), which represents the probability that our model will correctly predict a higher probability for a randomly chosen positive case than a randomly chosen negative case. We calculated Brier score, which is the mean squared difference between predicted probabilities and true outcomes. It can be represented as: $\text{Brier Score} = \frac{1}{n_i} \Sigma_{i=1}^{n} (\hat{p(X)_i^{(0)}}-Y_i^{(0)})^2$. We also calculated Brier skill score, which is a measure of how the model does relative to a model that predicts the mean predicted probability for every observation. All of these metrics measure the accuracy of individual predictions or individual predictions relative to each other. However, because attribution estimate is dependent on the accuracy of the mean probability, we propose a fourth metric: mean calibration error (MCE). This can be represented as $\text{MCE} =|\frac{1}{n}\Sigma_{i=1}^{n}\hat{p(X)_i^{(0)}}- \frac{1}{n}\Sigma_{i=1}^{n}Y_i^{(0)}|$, the absolute value of the difference between the mean predicted probability and the true proportion of outcomes. 

We evaluated the correlation between standard predictive performance metrics and average log risk ratio error across all 25 combinations of machine learning and calibration methods. We calculated this correlation for each metric in each of our 500 simulated datasets, taking the average correlation across the 500 simulations for our final analysis to account for the Bernoulli sampling variation in the creation of our datasets.

\subsubsection{Paired T-tests of differences in Fisher’s Z scores across performance metrics}
To assess whether the correlations between performance metric and average log risk ratio error differ significantly between performance metrics, we perform paired t-tests of the differences in the Fisher’s Z transformed correlations across all 500 simulated datasets. We compare the following metrics: AUC vs. Brier Skill Score, AUC vs. Mean Calibration Error, and Brier Skill Score vs. Mean Calibration Error.

\subsubsection{Regret analysis for performance metric accuracy}
To further assess the effectiveness of performance metrics at identifying the most accurate machine learning method, we perform a worst-case regret analysis. For a given performance metric, we rank all 25 candidate models using this metric and select the 5 models with the best performance. We then identify the highest absolute log risk ratio error across these 5 models. We repeat this for each of the 500 simulated datasets and compare the distributions of highest absolute log risk ratio error between performance metrics to assess which metrics best identify more accurate models.

\subsubsection{Distribution shift diagnostics}
To address concerns about distribution shift under anthropogenic climate change, we assess using methods of propensity score weighting. Propensity score in this context is the estimated probability of a fire day being from the present-day climate scenario based on our climate, land, and weather predictors. To estimate the propensity score, we create combined datasets of our predictors from the present-day scenario and counterfactual scenario and include a binary indicator variable for the source climate scenario (e.g. ``1” for present-day, ``0” for counterfactual). We create separate datasets for each counterfactual scenario. We use the same sample splitting method by three-year time period to predict the probability of a fire day being in the present-day scenario. We tuned our propensity models via cross-validation using LightGBM. The result is a set of propensity scores for the present-day and counterfactual datasets using out-of-sample predictions; the present-day scenario propensity scores change with each counterfactual scenario.

We chose LightGBM as the propensity model for its computational efficiency and ability to flexibly model nonlinear relationships amongst predictors; as robustness checks, we also tested a variety of different propensity models. 

We used principal components analysis to visualize and summarize distribution shifts across climate scenarios. We calculated proxy-A distance as a measure of distribution shift, which is represented as: $2 - 4e$, where $e = 1 - accuracy$. In this case, accuracy is calculated as the percentage of correctly labeled points. We used a positive label threshold of 0.5.\autocite{ben-david_analysis_2006}

To assess the impact of distribution shift on predictive performance, we performed a temperature subanalysis on our present-day data. We binned our present-day data into eight groups by temperature and recalculated mean calibration error in each bin. 

We calculate out-of-sample mean calibration error on simulated datasets with known true probabilities to further assess the impact of distribution shift.  Out-of-sample mean calibration error is calculated as $MCE =|\frac{1}{n}\Sigma_{i=1}^{n}\hat{p(X)_i^{(1)}}- \frac{1}{n}\Sigma_{i=1}^{n}p(X)_i^{(1)}|$.

\subsubsection{Evaluating Prediction-Powered Inference via simulation}
To address the concern that using the average of our predictions assumes our predictions to be perfect and fails to capture variance due to imperfect machine learning predictions, we evaluate the use of Prediction-Powered Inference (PPI) using simulation.\autocite{angelopoulos_prediction-powered_2023} Prediction-Powered Inference uses information from the performance of the machine learning model on our labeled, observed data to inform the estimate of mean probability in our unlabeled counterfactual dataset. 

We generate our true probabilities of extreme daily growth ($p(X)^{(0)}$ and $p(X)^{(1)}$) using the following equation:
\[logit(p(X)) = \beta_0+\alpha h_d(X) +\sigma \epsilon\]
where $h_d(X)$ is a fixed, randomly-generated nonlinear risk across predictor variables $X$ and draw $d$ (Supplement), $\epsilon$ is noise, and $\beta_0$ is the intercept that ensures the probability of a extreme fire growth in the simulated present-day dataset matches the observed present-day dataset.\autocite{dorie_automated_2019} $h_d(X)$ is calculated as a combination of a linear component and a random Fourier feature approximation (a standard Rahimi \& Recht approximation to the Gaussian kernel)\autocite{rahimi_random_2007} using a cosine basis function (Supplement). $\alpha$ and $\sigma$ are a function of $\rho$, where $\rho$ represents variance attributable to the observed predictors:
\[\alpha = \tau \sqrt{\rho}\]
\[\sigma = \tau \sqrt{1-\rho}\]
where $\tau = 3.5$, representing the total logit scale spread.

We generate 50 $h_d(X)$ equations, representing 50 different data generating processes. For each $h_d(X)$, we generate 20 simulated present-day datasets where variation sources from Bernoulli sampling of the response variable, $Y$, and noise variable, $\epsilon$, and 20 corresponding simulated datasets for each counterfactual climate scenario (pre-industrial and SSP5-8.5, end-of-century) where variation sources from the noise variable. We repeat this process for three values of $\rho$ -- 0.6, 0.8, 0.95 -- for a total of 6000 simulations.

We train and tune LightGBM models on each simulated present-day dataset using cross validation to predict the probability of extreme daily growth in the corresponding simulated pre-industrial and SSP5-8.5 scenarios. We make attribution estimates using three different methods of generating $E[Y^{(1)}]$ from these predicted probabilities: plug-in, unweighted PPI, and importance weighted PPI.

We generate 95\% confidence intervals for each of our attribution estimates. For estimates calculated using the plug-in method, we use the delta method calculation of the confidence interval (Supplement). For estimates calculated using PPI, we use a bootstrap method similar to the method described in section 7.2.4. We sample 17,910 days with replacement in the simulated present-day dataset and their corresponding counterfactual scenario days, retrain LightGBM models, refit importance weights (if weighted PPI), and estimate RR using PPI.  We repeat this method 100 times, taking the 2.5 and 97.5 percentile estimates to get the 95\% confidence interval. 

We use log risk ratio error as a measure of attribution accuracy, which can be represented as: \[\log{(RR_{error)}} = \log{\frac{\hat{Y^{(0)}}}{\hat{p(X)^{(1)}}}} - \log{\frac{p(X)^{(0)}}{p(X)^{(1)}}}\]. 

We also calculate coverage, which is the percentage of confidence intervals for each method ($\rho$, attribution method, climate scenario) that contain the true risk ratio.

\begin{appendices}
\newpage
\section*{Supplementary Information}
\setcounter{table}{0}  
\renewcommand{\thetable}{S\arabic{table}}
\setcounter{figure}{0}
\renewcommand{\thefigure}{S\arabic{figure}}

\begin{table}[ht]
    \centering
    \caption[Aggregate risk ratio estimates (95\% Confidence Interval)]{Aggregate risk ratio estimates, weighted PPI method (95\% Confidence Interval)}
    \begin{tabular}{lcc}
    \toprule
    \textbf{Machine Learning} & \textbf{Pre-Industrial vs.} & \textbf{Present-Day vs. } \\
    \textbf{Method} & \textbf{Present-Day} & \textbf{SSP5-8.5, end-of-century} \\
    \midrule
        LightGBM & 1.18 (1.13, 1.25) & 1.79 (1.47, 2.09) \\
        Random Forest & 1.21 (1.16, 1.27) & 2.20 (1.88, 2.54) \\
        XGBoost & 1.27 (1.06, 1.36) & 1.82 (1.43, 2.11) \\
        Logistic Regression & 1.24 (1.18, 1.30) & 1.83 (1.56, 2.15) \\
        Elastic Net Regression & 1.24 (1.19, 1.31) & 1.90 (1.56, 2.24) \\
        Ensemble (Superlearner) & 1.21 (1.11, 1.37) & 2.09 (1.48, 2.44) \\
        \bottomrule
    \end{tabular}
    \label{tab:supp1}
\end{table}

\newpage
\begin{table}[ht]
   \centering
    \caption{Paired-T Tests of difference in Fisher’s Z scores, a transformation of Pearson's correlation coefficient of predictive performance and absolute log risk ratio error, between different predictive performance metrics across 500 simulations.}
    \begin{tabular}{lccccc}
        \toprule
        \multicolumn{6}{c}{\textbf{Pre-Industrial vs. Present Day}} \\
        \midrule
        \textbf{Metrics} & \textbf{Mean (95\% CI)} & \textbf{Standard Error} & \textbf{t} & \textbf{P-Value} & \textbf{df} \\
        \midrule
        AUC - BSS & -0.292 (-0.319, -0.265) & 0.014 & -21.25 & $<$ 0.0001 & 499 \\
        AUC - MCE & -0.737 (-0.779, -0.696) & 0.021 & -34.97 & $<$ 0.0001 & 499 \\
        BSS - MCE & -0.445 (-0.483, -0.407) & 0.019 & -23.17 & $<$ 0.0001 & 499 \\
        \midrule
        \multicolumn{6}{c}{\textbf{Present Day vs. SSP5-8.5, End-of-Century}} \\
        \midrule
        AUC - BSS & -0.0004 (-0.034, 0.032) & 0.017 & -0.026 & 0.980 & 499 \\
        AUC - MCE & 0.005 (-0.024, 0.034) & 0.015 & 0.345 & 0.730 & 499 \\
        BSS - MCE & 0.006 (-0.017, 0.028) & 0.011 & 0.485 & 0.628 & 499 \\
        \bottomrule
    \end{tabular}
    \label{tab:supp2}
\end{table}

\newpage
\begin{table}[ht]
    \centering
    \caption[Regret demonstrates that predictive performance decreases with distribution shift.]{Regret demonstrates that predictive performance decreases with distribution shift. Median regret, calculated as the highest absolute log risk ratio error across the 5 models with the best in-sample performance metric, across 7500 simulations.}
    \begin{tabular}{lccc}
        \toprule
        \textbf{Scenario} & \textbf{AUC} & \textbf{Brier Skill Score} & \textbf{Mean Calibration Error} \\
        \midrule
        Pre-Industrial & 0.193 & 0.110 & 0.134 \\
        SSP5-8.5, End of Century & 0.249 & 0.249 & 0.359 \\
        \bottomrule
    \end{tabular}
    \label{tab:supp3}
\end{table}

\newpage
\begin{table}[ht]
    \centering
    \caption[Post-hoc calibration does not improve predictive performance.]{Post-hoc calibration does not improve predictive performance. Predictive performance metrics of calibrated predictions from observed data.}
    \begin{tabular}{lcccc}
        \toprule
        \textbf{ML Method} &  \textbf{Calibration Method} & \textbf{AUC} & \textbf{Brier Skill Score} & \textbf{Mean Calibration Error} \\
        \midrule
        LightGBM & None & 0.890 & 0.041 & 0.0035 \\
        LightGBM & Logistic & 0.870 & 0.042 & 0.0006 \\
        LightGBM & Isotonic & 0.872 & -0.006 & 0.0081 \\
        LightGBM & Natural Spline & 0.863 & 0.044 & 0.0007 \\
        LightGBM & I-Spline & 0.868 & 0.036 & 0.0009 \\
        Random Forest & None & 0.879 & 0.066 & 0.0021 \\
        Random Forest & Logistic & 0.858 & 0.054 & 0.0005 \\
        Random Forest & Isotonic & 0.858 & 0.032 & 0.0070 \\
        Random Forest & Natural Spline & 0.858 & 0.055 & 0.0006 \\
        Random Forest & I-Spline & 0.863 & 0.046 & 0.0003 \\
        XGBoost & None & 0.864 & 0.043 & 0.0017 \\
        XGBoost & Logistic & 0.838 & 0.031 & 0.0004 \\
        XGBoost & Isotonic & 0.843 & -0.029 & 0.0124 \\
        XGBoost & Natural Spline & 0.839 & 0.033 & 0.0007 \\
        XGBoost & I-Spline & 0.842 & 0.026 & 0.0005 \\
        Logistic Regression & None & 0.877 & 0.041 & 0.00006 \\
        Logistic Regression & Logistic & 0.852 & 0.035 & 0.0005 \\
        Logistic Regression & Isotonic & 0.852 & -0.019 & 0.0074\\
        Logistic Regression & Natural Spline & 0.853 & 0.035 & 0.0005 \\
        Logistic Regression & I-Spline & 0.846 & 0.022 & 0.0006 \\
        Elastic Net & None & 0.865 & 0.041 & 0.0004 \\
        Elastic Net & Logistic & 0.841 & 0.033 & 0.0005 \\
        Elastic Net & Isotonic & 0.838 & -0.088 & 0.0087\\
        Elastic Net & Natural Spline & 0.834 & 0.029 & 0.0005 \\
        Elastic Net & I-Spline & 0.839 & 0.013 & 0.0005 \\
        \bottomrule
    \end{tabular}
    \label{tab:supp4}
\end{table}

\newpage
\begin{table}[ht]
    \centering
    \caption[Aggregate fraction of attributable risk estimates, plug-in method (95\% Confidence Interval)]{Aggregate fraction of attributable risk estimates, plug-in method (95\% Confidence Interval)}
    \begin{tabular}{lcc}
    \toprule
    \textbf{Machine Learning} & \textbf{Pre-Industrial vs.} & \textbf{Present-Day vs. } \\
    \textbf{Method} & \textbf{Present-Day} & \textbf{SSP5-8.5, end-of-century} \\
    \midrule
        LightGBM & 0.156 (0.114, 0.208) & 0.313 (0.146, 0.446) \\
        Random Forest & 0.192 (0.142, 0.228) & 0.583 (0.558, 0.670) \\
        XGBoost & 0.286 (-0.175, 0.370) & 0.477 (0.241, 0.550) \\
        Logistic Regression & 0.172 (0.128, 0.214) & 0.529 (0.467, 0.588) \\
        Elastic Net Regression & 0.192 (0.145, 0.224) & 0.584 (0.491, 0.628) \\
        Ensemble (Superlearner) & 0.170 (0.141, 0.312) & 0.58 (0.459, 0.633) \\
        \bottomrule
    \end{tabular}
    \label{tab:tab1}
\end{table}

\newpage
\begin{table}[ht]
    \centering
    \caption[Aggregate risk ratio estimates, plug-in method (95\% Confidence Interval)]{Aggregate risk ratio estimates, plug-in method (95\% Confidence Interval)}
    \begin{tabular}{lcc}
    \toprule
    \textbf{Machine Learning} & \textbf{Pre-Industrial vs.} & \textbf{Present-Day vs. } \\
    \textbf{Method} & \textbf{Present-Day} & \textbf{SSP5-8.5, end-of-century} \\
    \midrule
        LightGBM & 1.18 (1.13, 1.26) & 1.46 (1.17, 1.81) \\
        Random Forest & 1.24 (1.17, 1.30) & 2.40 (2.26, 3.03) \\
        XGBoost & 1.40 (0.85, 1.59) & 1.91 (1.32, 2.22) \\
        Logistic Regression & 1.21 (1.15, 1.27) & 2.12 (1.88, 2.43) \\
        Elastic Net Regression & 1.24 (1.17, 1.29) & 2.40 (1.96, 2.69) \\
        Ensemble (Superlearner) & 1.21 (1.16, 1.45) & 2.38 (1.85, 2.73) \\
        \bottomrule
    \end{tabular}
    \label{tab:supp1}
\end{table}

\newpage
\begin{figure}[ht]
    \centering
    \includegraphics[width=1\linewidth]{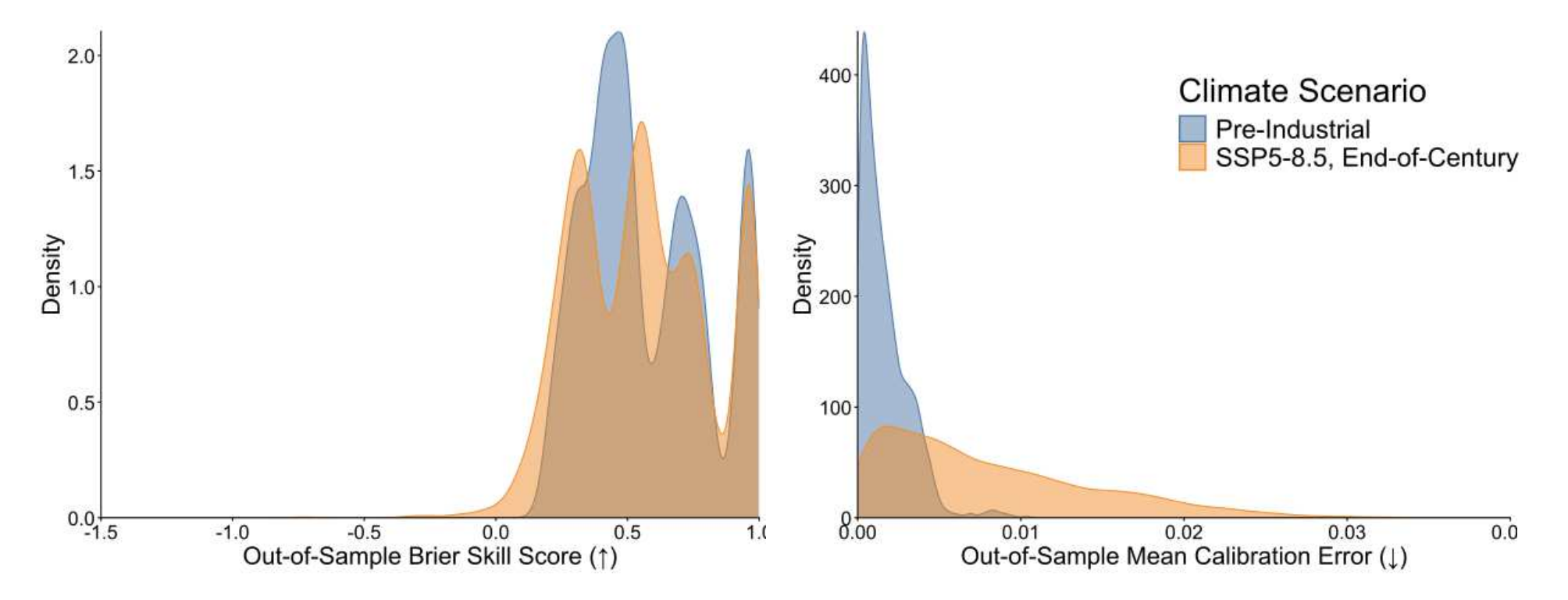}
    \caption[Out-of-sample performance metrics perform better in the pre-industrial climate scenario than in the SSP5-8.5, end-of-century scenario.]{\textbf{Out-of-sample performance metrics perform better in the pre-industrial climate scenario than in the SSP5-8.5, end-of-century scenario.}}
    \label{fig:suppfig2}
\end{figure}

\newpage
\begin{figure}[ht]
    \centering
    \includegraphics[width=1\linewidth]{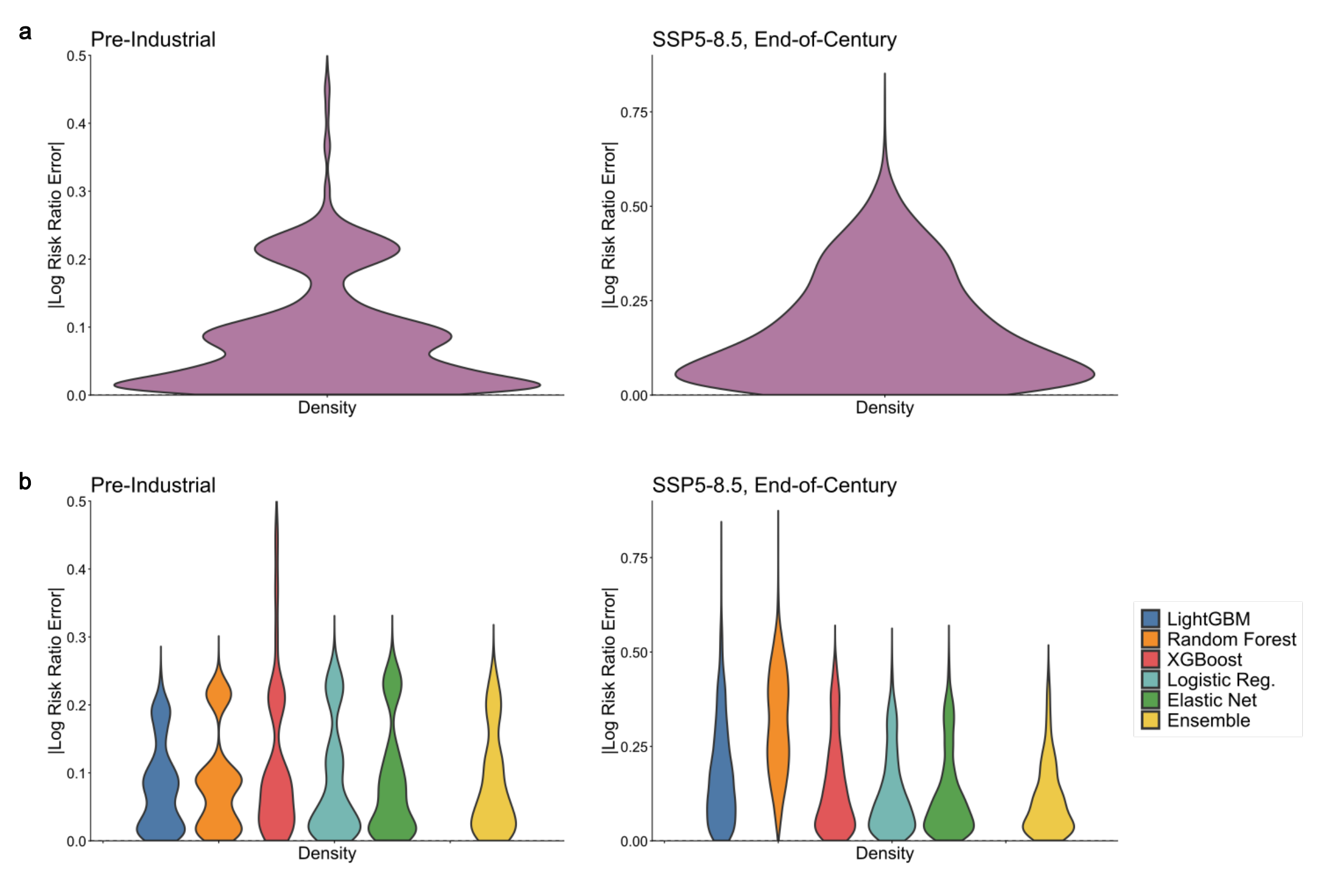}
    \caption[Absolute value of log risk ratio error, by model.]{\textbf{Absolute value of log risk ratio error, by model.} \textbf{a}, Distribution of the absolute value of log risk ratio error estimates from 9000 simulations. Individual points represent log risk ratio error estimates from approximately 15 simulations. Underlying grey violin plot indicates density across all models in the simulations. \textbf{b}, Distribution of the absolute value of log risk ratio error estimates from 1500 simulations for each of the 5 machine learning models.}
    \label{fig:suppfig1}
\end{figure}

\newpage
\section{Supplementary Note: Methods}
\label{appendix:partA}
\subsection{Estimating Individual Fire Day Fraction of Attributable Risk ($FAR_i$)}
To introduce notation, let $Y$ be the vector length $n$ of observed binary outcomes of extreme fire growth for $n$ fire days. Let $X$ represent the $n$ x $d$ matrix of $d$ predictors for $n$ fire days. Let $p(X)$ represent the vector length $n$ of predicted probabilities of extreme fire growth for $n$ fire days.

For an individual fire day, the fraction of attributable risk is calculated as:
\[ FAR_i =1-\frac{E[Y_i^{(0)}|X_i^{(0)}]}{E[Y_i^{(1)}|X_i^{(1)}]}\]
where $E[Y_i^{(0)} | X_i^{(0)}]$  is the probability of extreme daily growth given predictor values on day $i$ in the real world and $E[Y_i^{(1)} | X_i^{(1)}]$ is the individual fire day probability of extreme daily growth given predictor values on day $i$ in the counterfactual world. Since both quantities are unobserved, we substitute with $\hat{p(X)_i^{(0)}}$ and $\hat{p(X)_i^{(1)}}$, our machine learning model estimates:
\[ \hat{FAR_i} =1-\frac{\hat{p(X)_i^{(0)}}}{\hat{p(X)_i^{(1)}}}\]

Similarly, the risk ratio (RR) and average treatment effect (ATE) can be represented as:
\[ RR_i =\frac{E[Y_i^{(1)}|X_i^{(1)}]}{E[Y_i^{(0)}|X_i^{(0)}]}\]
\[ \hat{RR_i} =\frac{\hat{p(X)_i^{(1)}}}{\hat{p(X)_i^{(0)}}}\]
\[ ATE_i=E[Y_i^{(1)}|X_i^{(1)}]-E[Y_i^{(0)}|X_i^{(0)}]\]
\[ \hat{ATE_i} = \hat{p(X)_i^{(1)}} - \hat{p(X)_i^{(0)}}\]

We note that when calculating attribution estimates, we treat $X_0, Y_0$ and $p(X)^{(0)}$ as the covariates, observed outcome, and probability in the historical (relative) climate scenario, and $X_1, Y_1$, and $p(X)^{(1)}$ as these values in the future (relative) climate scenario. As such, when comparing the Present-Day and Pre-Industrial climate scenarios, $X_0$ and $p(X)^{(0)}$ represents the pre-industrial values ($Y_0$ is unobserved) and $X_1, Y_1$, and $p(X)^{(1)}$ represent the present-day values. Similarly, when comparing the Present-Day and SSP5-8.5, end of century climate scenarios, $X_0, Y_0$, and $p(X)^{(0)}$ represent the present-day values, and $X_1$ and $p(X)^{(1)}$ represent the SSP5-8.5, end of century values ($Y_1$ is unobserved).

\subsection{Estimating Aggregate Fraction of Attributable Risk (FAR)}
FAR, RR, and ATE can also be calculated as aggregate quantities that represent the average attribution of changes in extreme fire growth to anthropogenic emissions. We focus on these aggregate attribution estimates through most of this analysis. They can be represented as:
\[ FAR =1-\frac{E[Y^{(0)}]}{E[Y^{(1)}]}\]
\[ RR =\frac{E[Y^{(1)}]}{E[Y^{(0)}]}\]
\[ ATE = E[Y^{(1)}] - E[Y^{(0)}]\]
The fire-day average probability of extreme daily growth in the real world, $E[Y^{(0)}]$ is calculated as the fraction of extreme daily growth days in our observed dataset: 380/17910. Because $E[Y^{(1)}]$ is unobserved, we calculate the expected probability in the counterfactual world as the average of the predicted probabilities in the counterfactual world, $E[\hat{p(X)^{(1)}}]= \frac{1}{n} \Sigma_{i=1}^{n} \hat{p(X)_i^{(1)}}$. This can be represented as:
\[ \hat{FAR} =1-\frac{E[Y^{(0)}]}{E[\hat{p(X)^{(1)}}]}\]
\[ \hat{RR} =\frac{E[\hat{p(X)^{(1)}}]}{E[Y^{(0)}]}\]
\[ \hat{ATE} = E[\hat{p(X)^{(1)}}] - E[Y^{(0)}]\]
We explored other methods of calculating expected probability in the counterfactual world, including Prediction-Powered Inference (PPI) and Prediction-Powered Inference using importance weighting (see appendix \ref{appendix:partA}).

\subsection{Assessing accuracy of FAR estimates}
Formally, let days be indexed by \(t=1,\dots,T\), methods by \(m=1,\dots,5\), and simulation replicates by \(r=1,\dots,R\) with \(R=300\). For each method \(m\), we define the “true” probabilities
\[
\pi_{t}^{(0,m)}\in[0,1]\quad\text{and}\quad \pi_{t}^{(1,m)}\in[0,1],
\]
obtained from our real- and counterfactual-world predicted probabilities, respectively. Let
\[
\mu_{0}^{(m)}=\frac{1}{T}\sum_{t=1}^{T}\pi_{t}^{(0,m)},\qquad
\mu_{1}^{(m)}=\frac{1}{T}\sum_{t=1}^{T}\pi_{t}^{(1,m)},
\]
and define the truth for the estimands as
\[
\mathrm{ATE}^{\star,(m)}=\mu_{1}^{(m)}-\mu_{0}^{(m)},\qquad
\mathrm{FAR}^{\star,(m)}=1-\frac{\mu_{0}^{(m)}}{\mu_{1}^{(m)}}.
\]

For each replicate \(r\) and method \(m\), we generate outcomes on the real-world design by
\[
Y_{t,r}^{(0,m)} \sim \mathrm{Bernoulli}\!\big(\pi_{t}^{(0,m)}\big)\quad (t=1,\dots,T),
\]
keeping the predictors \(X^{(0)}\) and \(X^{(1)}\) fixed across replicates. We then train the model on \(\mathcal D_{0,r}^{(m)}=(X^{(0)},Y_{r}^{(0,m)})\) and apply it to \(X^{(1)}\) to obtain
\[
\hat p_{t,r}^{(1,m)}=\hat f_{r}^{(m)}\!\big(x_{t}^{(1)}\big).
\]

From there, we are able to compare our predicted probabilities $\hat p_{t,r}^{(1,m)}$ to the ``true" probabilities to calculate attribution error.

\subsection{Paired T-tests of differences in Fisher’s Z scores across performance metrics}
We transformed correlations into normally distributed Fisher’s Z values through the following equation:
\[ z = 0.5 \ln{\frac{1+r}{1-r}}\]
where $r$ is Pearson's correlation coefficient and $z$ is the Fisher’s Z score.

\subsection{Prediction-Powered Inference (PPI)}
Prediction-Powered Inference corrects potential machine learning bias through the addition of a rectifier, which is a measure of prediction error.\autocite{angelopoulos_prediction-powered_2023} The rectifier is computed by calculating the difference in the most appropriate measure of fit on the labeled data with the true outcomes vs. the model’s predictions on the labeled data. PPI does not make assumptions on the machine learning model used or the shape of the data distribution. PPI does assume that the labeled set and the unlabeled set are drawn from the same underlying distribution. If there is covariate shift, one can reweight the data to continue using PPI for valid inference. One method of reweighting includes importance weighting – adjusting for differences in covariate distributions such that the covariate structure of the small labeled set resembles the unlabeled set.  

\subsection{Generation of relationship between covariates and true probability, $h_d(X)$ in Prediction-Powered Inference simulations}
We follow a similar simulation setup to previous work by Dorie et al.\autocite{dorie_automated_2019} $h_d(X)$ is a random function calculated as a combination of a linear and a nonlinear component. The linear component is calculated as 
\[l(X)=X_{standardized}\beta_l\] where $\beta_l$ is drawn from $N(0,1)$ and normalized to unit length. The nonlinear component is calculated as:
\[ nl(X)=\Sigma_{k=1}^{n_r}w_{n}(\sqrt{\frac{2}{n}}\cos(X_{standardized}*\Omega_k+b_k))\]
where $n$ is the number of is the number of random Fourier features (a standard Rahimi \& Recht approximation to the Gaussian kernel\autocite{rahimi_random_2007}), $w_{n}$ is the vector of weights for each feature drawn from the standard normal distribution, $N(0, 1)$, $\Omega_k$ is a random frequency drawn from $N(0, \frac{1}{\text{length scale}^2})$, and $b_k$ is a random phase drawn from $Uniform(0, 2\pi)$.
We standardize and combine these components such that:
\[h_{raw}(X)= \sqrt{\gamma}*z(l(X))+\sqrt{1-\gamma}*z(nl(X))\]
\[h_{d}(X)=z(h_{raw}(X))\]
where $\gamma$ controls the weight of the linear component. This function is determined using the present-day dataset and applied to the counterfactual datasets.

\subsection{Analytical calculation of 95\% confidence interval in PPI simulations}
We calculate confidence intervals for the plug-in attribution estimates using the delta method of calculating standard error and the Wald interval. We represent the variance of the mean present-day outcome, $E[\hat{{Y^{(0)}}}]$, as:
\[ \hat{Var}(E[\hat{{Y^{(0)}}}]) = \frac{E[\hat{{Y^{(0)}}}](1-E[\hat{{Y^{(0)}}}])}{n}\]
We represent the variance of the mean predicted probabilities, $E[\hat{{p(X)^{(1)}}}]$, as:
\[ \hat{Var}(E[\hat{{p(X)^{(1)}}}]) = \frac{SD_{\hat{{p(X)^{(1)}}}}^2}{n}\]
where
\[ SD_{\hat{{p(X)^{(1)}}}}^2 = \frac{1}{n-1}\Sigma(\hat{{p(X)^{(1)}}}-E[\hat{{p(X)^{(1)}}}])^2\]
Using delta method, we calculate:
\[\hat{Var}(logRR)\approx\frac{\hat{Var}(E[\hat{{p(X)^{(1)}}}]}{E[\hat{{p(X)^{(1)}}}]^2}+\frac{\hat{Var}(E[\hat{{Y^{(0)}}}])}{E[\hat{{Y^{(0)}}}]^2}\]
We use $\hat{Var}(logRR)$ to calculate the Wald interval for logRR:
\[CI_{95\%}=\hat{logRR} \pm 1.96 \sqrt{\hat{Var}(logRR)} \]

\end{appendices}

\end{document}